\pdfoutput=1

\documentclass[11pt,twoside,a4paper,cmspaper,final,collab]{cms-tdr}
\def\svnVersion{284725}\def\cmsCernNo{2013-037}\def\cmsCernDate{\today}\def\cmsMessage{Published in the Journal of High Energy Physics as \href{http://dx.doi.org/10.1007/JHEP03(2015)022}{\doi{10.1007/JHEP03(2015)022.}}}

\begin{document}\cmsNoteHeader{HIN-13-004}

\hyphenation{had-ron-i-za-tion}
\hyphenation{cal-or-i-me-ter}
\hyphenation{de-vices}

\RCS$Revision: 273397 $
\RCS$HeadURL: svn+ssh://svn.cern.ch/reps/tdr2/papers/HIN-13-004/trunk/HIN-13-004.tex $
\RCS$Id: HIN-13-004.tex 273397 2015-01-12 10:29:24Z lamia $
\providecommand{\sqrtsnn}{\ensuremath{\sqrt{s_\mathrm{NN}}}\xspace}
\providecommand{\NA}{---}
\newcommand{\mbinv} {\ensuremath{\,\text{mb}^{-1}}\xspace}
\cmsNoteHeader{HIN-13-004} 
\title{Study of Z production in PbPb and pp collisions at \texorpdfstring{$\sqrtsnn = 2.76$\TeV}{sqrt(s[NN]) = 2.76 TeV} in the dimuon and dielectron decay channels}

\date{\today}

\abstract{The production of Z bosons is studied in the dimuon and dielectron decay channels in PbPb and pp collisions at $\sqrtsnn = 2.76$\TeV, using data collected by the CMS experiment at the LHC. The PbPb data sample corresponds to an integrated luminosity of about 166\mubinv, while the pp data sample collected in 2013 at the same nucleon-nucleon centre-of-mass energy has an integrated luminosity of 5.4\pbinv. The Z boson yield is measured as a function of rapidity, transverse momentum, and collision centrality. The ratio of PbPb to pp yields, scaled by the number of inelastic nucleon-nucleon collisions, is found to be $1.06\pm0.05\stat\pm0.08\syst$ in the dimuon channel and $1.02\pm0.08\stat\pm0.15\syst$ in the dielectron channel, for centrality-integrated Z boson production. This binary collision scaling is seen to hold in the entire kinematic region studied, as expected for a colourless probe that is unaffected by the hot and dense QCD medium produced in heavy ion collisions.
}

\hypersetup{%
pdfauthor={CMS Collaboration},%
pdftitle={Study of Z production in PbPb and pp collisions at sqrt(s[NN]) = 2.76 TeV in the dimuon and dielectron decay channels},%
pdfsubject={CMS},%
pdfkeywords={CMS, physics, heavy ions, Z boson}}

\maketitle

\section{Introduction}
The Z boson was first observed by the UA1 and UA2 experiments at CERN in proton-antiproton collisions at a centre-of-mass energy of 540\GeV~\cite{Arnison:1983mk,Bagnaia:1983zx}. Since then, its properties have been characterized in detail by a succession of collider experiments~\cite{Schael:2013ita, ALEPH:2005ab, Acosta:2004uq, Abachi:1995xc, CMS:2011aa, Chatrchyan:2011wt, Chatrchyan:2014mua, Aad:2011dm}. These properties, including mass and decay widths, as well as inclusive and differential cross sections, have been measured at different centre-of-mass energies in electron-positron, proton-proton, and proton-antiproton collisions. The large centre-of-mass energy and substantial integrated luminosities delivered by the CERN LHC for Pb beams provide new opportunities to study Z boson production in nucleus-nucleus collisions.

The Z bosons decay with a typical lifetime of 0.1\unit{fm/$c$} and their leptonic decays are of particular interest since leptons pass through the medium being probed without interacting strongly.
Dileptons from Z boson decays can thus serve as a control for the processes expected to be heavily modified in the hot and dense medium, such as quarkonium or Z+jet production~\cite{NoInteractingZ}. However, in heavy ion collisions, Z boson production can be affected by initial-state effects. The modification of the yield in heavy ion collisions is expected to be about 3\% from isospin effects, the result of the proton-neutron (or u-d quark) ratio being different in protons and Pb nuclei~\cite{Isospin}, and from multiple scattering and energy loss of the initial partons~\cite{NRJloss}. In addition, the nuclear modification of parton distribution functions (PDF) can lead to rapidity-dependent changes on the order of up to 5\% in the observed Z boson yield in PbPb collisions~\cite{Isospin}.

Based on the first PbPb collisions at the LHC, with an integrated luminosity of about 7\mubinv, the CMS collaboration reported results on the $\cPZ\to\Pgmp\Pgmm$~\cite{Chatrchyan:2011ua}, $\PW^\pm\to\mu^\pm\nu$~\cite{Chatrchyan:2012nt} and isolated photon~\cite{Chatrchyan:2012vq} production. These measurements show that electroweak bosons are essentially unmodified by the hot and dense medium. In this paper, Z boson production at $\sqrtsnn=2.76$\TeV is studied using PbPb collision data collected in 2011, which corresponds to an integrated luminosity of about 166\mubinv. From this PbPb data-taking period, the ATLAS collaboration published Z boson yields showing no deviation with respect to theoretical predictions~\cite{Aad:2012ew}. The set of pp data at the same centre-of-mass energy recorded in 2013 by CMS, with a total integrated luminosity of 5.4\pbinv, is used to measure pp yields. The nuclear modification factor ($R_\mathrm{AA}$), the ratio of PbPb and pp yields scaled by the number of inelastic nucleon-nucleon binary collisions, is then calculated. These larger PbPb and pp data samples allow a more precise measurement of the Z boson yield dependence on transverse momentum (\pt), rapidity ($y$), and collision centrality. The dimuon channel is analyzed with an improved reconstruction algorithm (described in Section~\ref{sec:Lep_reco}) compared to Ref.~\cite{Chatrchyan:2011ua}, and the electron channel is used for the first time in CMS to measure Z boson production in heavy ion collisions.

\section{The CMS detector}

The central feature of the CMS apparatus is a superconducting solenoid of 6\unit{m} internal diameter, providing a magnetic field of 3.8\unit{T}. Within the superconducting solenoid volume are a silicon pixel and strip tracker, a lead tungstate crystal electromagnetic calorimeter (ECAL), and a brass/scintillator hadron calorimeter (HCAL), each composed of a barrel and two endcap sections. Muons are measured in gas-ionization detectors embedded in the steel flux-return yoke outside the solenoid. Extensive forward calorimetry complements the coverage provided by the barrel and endcap detectors. A more detailed description of the CMS detector, together with a definition of the coordinate system used and the relevant kinematic variables, can be found in Ref.~\cite{Chatrchyan:2008zzk}.

Muons are measured in the pseudorapidity range $\abs{\eta}<2.4$ using three technologies: drift tubes, cathode strip chambers, and resistive-plate chambers. Matching muons to tracks measured in the silicon tracker results in a relative transverse momentum resolution for muons with $20 <\pt < 100\GeV$ of 1.3--2.0\% in the barrel and better than 6\% in the endcaps. The \pt resolution in the barrel is better than 10\% for muons with \pt up to 1\TeV~\cite{Chatrchyan:2012xi}. Electrons are measured in the ECAL that consists of 75\,848 lead tungstate crystals providing a pseudorapidity coverage in the barrel region (EB) of $\abs{\eta}<1.48$  and in the two endcap regions (EE) of $1.48<\abs{\eta}<3.0$. The ECAL energy resolution for electrons with a transverse energy $\ET\approx45$\GeV, which is typical of $\cPZ\to\Pep\Pem$ decays, is better than 2\% in the central region of the ECAL barrel $(\abs{\eta}<0.8)$, and is between 2\% and 5\% elsewhere. For low-bremsstrahlung electrons, where 94\% or more of their energy is contained within a $3\times3$ array of crystals, the energy resolution improves to 1.5\% for $\abs{\eta}<0.8$~\cite{Chatrchyan:2013dga}. Matching ECAL clusters to tracks measured in the silicon tracker is used to differentiate electrons from photons.
Two steel/quartz-fiber Cherenkov hadron forward calorimeters (HF) are used to estimate the centrality of the PbPb collisions. The HF detectors are located on each side of the interaction point, covering the pseudorapidity region $2.9<\abs{\eta}<5.2$.

\section{Event selection and centrality determination}
\label{sec:selection}

In order to select a sample of purely inelastic hadronic PbPb collisions, the contamination from ultraperipheral collisions and non-collision beam background is removed, as described in Ref.~\cite{CMSJet}. Events are preselected if they contain a reconstructed primary vertex containing at least two tracks and at least three HF towers on each side of the interaction point with an energy of at least 3\GeV deposited in each tower. To further suppress the beam-gas events, the distribution of hits in the pixel detector along the beam direction is required to be compatible with particles originating from the event vertex. These criteria select $(97\pm3)\%$ of hadronic PbPb collisions~\cite{CMSJet}, corresponding to a number of efficiency-corrected minimum bias (MB) events $N_\mathrm{MB}=(1.16\pm0.04)\times10^9$ for the sample analyzed. 
The pp data set corresponds to an integrated luminosity of 5.4\pbinv known to an accuracy of 3.7\% from the uncertainty in the calibration based on a van der Meer scan~\cite{CMS-PAS-LUM-13-002}.

For the $\cPZ\to \Pgmp\Pgmm$ study in PbPb collisions, a trigger requiring a single muon with \pt greater than 15\GeVc is used, while a double-muon trigger with no explicit \pt selection is used for the pp sample. The efficiency for triggering on the $\cPZ\to\Pgmp\Pgmm$ channel within the selection and the acceptance requirements of the analysis is approximately 99\% and 98\% in the case of PbPb and pp collisions, respectively. For the $\cPZ\to\Pep\Pem$ channel in both PbPb and pp collisions, a trigger requires two significant energy deposits in the ECAL, one with $\ET>20$\GeV and another with $\ET>15$\GeV. The trigger efficiencies for the $\cPZ\to\Pep\Pem$ channel are approximately 96\% and 99\% for Z bosons produced in PbPb and pp collisions, respectively. The difference comes from the energy-clustering algorithms used at the trigger level in PbPb~\cite{Ball:2007zza} and pp~\cite{Chatrchyan:2011ue} collisions.

Centrality for PbPb collisions is defined by the geometrical overlap of the incoming nuclei, and allows for splitting up the PbPb data into centrality classes ranging from peripheral, where there is little overlap of the colliding nuclei, to central, where there is nearly complete overlap of the colliding nuclei. In CMS, the centrality of a PbPb collision is defined through bins that correspond to fractions of the total hadronic inelastic cross section as observed in the distribution of the sum of the transverse energy deposited in the HF calorimeters~\cite{CMSJet}. The centrality classes used in this analysis are 50--100\% (most peripheral), 40--50\%, 30--40\%, 20--30\%, 10--20\%, and 0--10\% (most central), ordered from the lowest to the highest HF energy deposit.

When measuring the nuclear modification factor, $R_\mathrm{AA}$, as described in Section~\ref{sec:R_AA}, the corrected Z boson yields in PbPb collisions are compared to those in pp collisions, scaled by the nuclear overlap function, $T_\mathrm{AA}$~\cite{Miller:2007ri}. At a given centrality, $T_\mathrm{AA}$ can be interpreted as the NN-equivalent integrated luminosity per nucleus-nucleus (AA) collision, and $T_\mathrm{AA}$-normalized Z boson yields can thus be directly compared with the Z boson production cross sections in pp collisions. In units of mb$^{-1}$, the average $T_\mathrm{AA}$ goes from $0.47\pm0.07$ to $23.2\pm1.0$, from the peripheral 50--100\% to the central 0--10\% ranges. These numbers, as well as all centrality-related quantities summarized in Table~\ref{tab:glauber}, are computed using the Glauber model~\cite{Miller:2007ri, Alver:2008aq}. The same parameters are used as in Ref.~\cite{CMSJet}, namely standard parameters for the Woods-Saxon function that distributes the nucleons in the Pb nuclei, and a nucleon-nucleon inelastic cross section of $\sigma_\mathrm{NN}^\text{inel}=64\pm5$\unit{mb}, based on a fit to the existing data for total and elastic cross sections in proton-proton and proton-antiproton collisions~\cite{Beringer:1900zz}. It is to be noted that the PbPb hadronic cross section ($7.65\pm0.42$~barns) computed with this Glauber Monte Carlo simulation results in an integrated luminosity of $152 \pm9$ \mubinv compatible within $1.2\sigma$ with the integrated luminosity based on the van der Meer Scan which has been evaluated to be $166\pm8$ \mubinv. All the results presented in the paper have been obtained using the Glauber model and event counting that is equivalent to $152$ \mubinv expressed in terms of luminosity.

\begin{table}[h!]
\centering
\topcaption{The average numbers of participating nucleons ($N_\text{part}$), binary collisions ($N_\text{coll}$), and the nuclear overlap function ($T_\mathrm{AA}$), corresponding to the centrality ranges used in this analysis.\label{tab:glauber}}
\begin{tabular}{ccccc}
  Centrality & $\langle N_\text{part} \rangle$ & $\langle N_\text{coll} \rangle$ & $\langle T_\mathrm{AA} \rangle$ (mb$^{-1}$)\\ \hline
  $[50,100]$\% & $22 \pm 2$  & $30  \pm 5$ & $0.47 \pm 0.07$ \\
  $[40,50]$\%  & $86 \pm 4$  & $176 \pm 21$   & $2.75 \pm 0.30$ \\
  $[30,40]$\%  & $130 \pm 5$ & $326 \pm 34$   & $5.09 \pm 0.43$ \\
  $[20,30]$\%  & $187 \pm 4$ & $563 \pm 53$   & $8.80 \pm 0.58$ \\
  $[10,20]$\%  & $261 \pm 4$ & $927 \pm 81$   & $14.5 \pm 0.80$ \\
  $[0,10]$\%   & $355 \pm 3$ & $1484 \pm 120$ & $23.2 \pm 1.00$ \\ \hline
  $[0,100]$\%  & $113 \pm 3$ & $363 \pm 32$   & $5.67 \pm 0.32$ \\
\end{tabular}
\end{table}

\section{Lepton reconstruction}
\label{sec:Lep_reco}

Muons are reconstructed using a global fit to a track in the muon detectors matched to a track in the silicon tracker. The offline muon reconstruction algorithm used for the PbPb data has been significantly improved relative to that used for the previous measurement~\cite{Chatrchyan:2011ua}. The efficiency has been increased by running multiple iterations in the pattern recognition step. The single-muon reconstruction efficiency is thus increased from $\simeq$85\% to $\simeq$98\% for muons from the Z boson decays with $\pt^{\mu}>20$\GeVc, reaching the efficiency level of the algorithm used for pp collisions. Background muons from cosmic rays and heavy-quark semileptonic decays are rejected by requiring a set of selection criteria on each muon track. The criteria used are based on previous studies of the performance of the muon reconstruction~\cite{Chatrchyan:2012xi}. At least one muon detector hit is required to be included in the global-muon track fit, and segments in at least two muon detectors are required to be matched to the track in the silicon tracker. To ensure a good \pt measurement, at least four tracker layers with a hit are required, and the $\chi^2$ per number of degrees of freedom of the global-muon track fit is required to be less than 10. To further reject cosmic muons and muons from decays in flight, the track is required to have a hit from at least one pixel detector layer and a transverse (longitudinal) distance of closest approach of less than 0.2 (5.0) mm from the measured primary vertex position. The most stringent selection criterion is the requirement of hits from more than one muon detector being matched to the global-muon track. The efficiency of these requirements is $\approx$98\% for muons from the Z boson decays, and after applying these selections, the $\cPZ\to\Pgmp\Pgmm$ charge misidentification rate is less than 1\% in both PbPb and pp collisions.

The electron reconstruction method uses information from the pixel and strip tracker, and the ECAL. Electrons traversing the silicon tracker can emit bremsstrahlung photons and deposit energy in the ECAL with a significant spread in the azimuthal direction. An algorithm for creating \emph{superclusters}, which are clusters of $\ET$ deposits from particles passing through the ECAL, is used for estimating the proper energy of photons in the heavy-ion environment, as in Ref.~\cite{Chatrchyan:2012vq}. A dedicated algorithm is used to reconstruct electrons that takes into account the bremsstrahlung emissions~\cite{Adam:2003kg}. Track seeds in the pixel detector compatible with the superclusters are found and used to initiate the construction of particle trajectories in the inner tracker. The standard algorithms and identification criteria presented in Ref.~\cite{Chatrchyan:2013dga} are used for the pp sample, resulting in a reconstruction efficiency of about 98\%. For PbPb collisions, the electron reconstruction efficiency is about 85\% for $\pt^{\Pe}>20$\GeVc electrons from the Z boson decays because the track reconstruction efficiency optimized for high-multiplicity events is lower than for pp collisions. The same electron identification variables are used in PbPb and pp collisions, with more stringent selection criteria in the latter case in order to match the ones in the $\sqrt{s}=7$\TeV pp analyses~\cite{CMS:2011aa, Chatrchyan:2011wt}. The requirements used in the selection process that have been found to be the most effective in reducing the background (see Ref.~\cite{Chatrchyan:2013dga} for definition of the variables) are: the energy-momentum combination between the supercluster and the track, the variables measuring the $\eta$ and $\phi$ spatial matching between the track and the supercluster, the supercluster shower shape width, the hadronic leakage (the ratio of energy deposited in the HCAL and ECAL), and a transverse distance of closest approach from the measured primary vertex. These selection criteria reduce the single-electron efficiency by about 10\% (5\%) in PbPb (pp) collisions. After applying these criteria, the $\cPZ\to\Pep\Pem$ charge misidentification rate for PbPb (pp) is less than 8\% (4\%), and is described well by a prediction of the combinatorial background based on same-charge pairs.

\section{Signal extraction, corrections, and systematic uncertainties}
\subsection{Signal extraction}

The Z boson candidates in PbPb and pp collisions are selected by requiring opposite-charge lepton pairs and then choosing those in the 60--120\GeVcc invariant mass region, where both leptons fulfill the acceptance and quality requirements. The acceptance requirements for both muons in PbPb and pp analyses are $\pt^{\mu}>20$\GeVc, to suppress muons from background processes, e.g. punch-through hadrons~\cite{Chatrchyan:2012xi}, and $\abs{\eta^{\mu}}<2.4$ given by the acceptance of the muon detectors. Both electrons are required to have $\pt^\Pe>20$\GeVc to suppress electrons from background processes, and $\abs{\eta^\Pe}<1.44$ to restrict them to be within the EB, to take advantage of a higher electron reconstruction efficiency and a better resolution in this region. In both channels, the dileptons are chosen to be in the experimentally visible region in rapidity. The dimuon system rapidity is limited to $\abs{y}<2.0$, while the rapidity of the dielectron system is limited to $\abs{y}<1.44$.

\begin{figure}[ht]
  \begin{center}
  \includegraphics[width=0.45\textwidth]{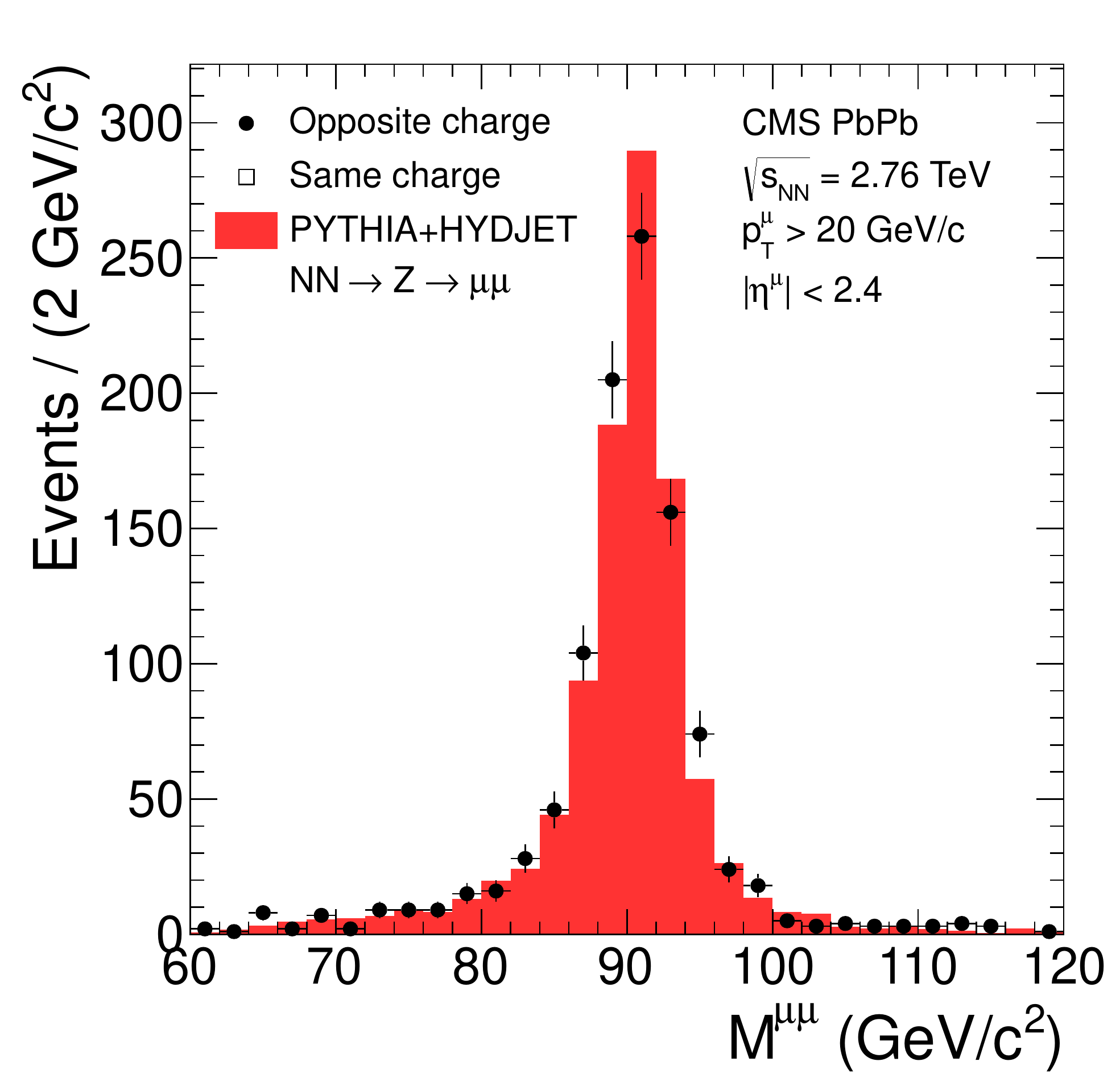}
  \includegraphics[width=0.45\textwidth]{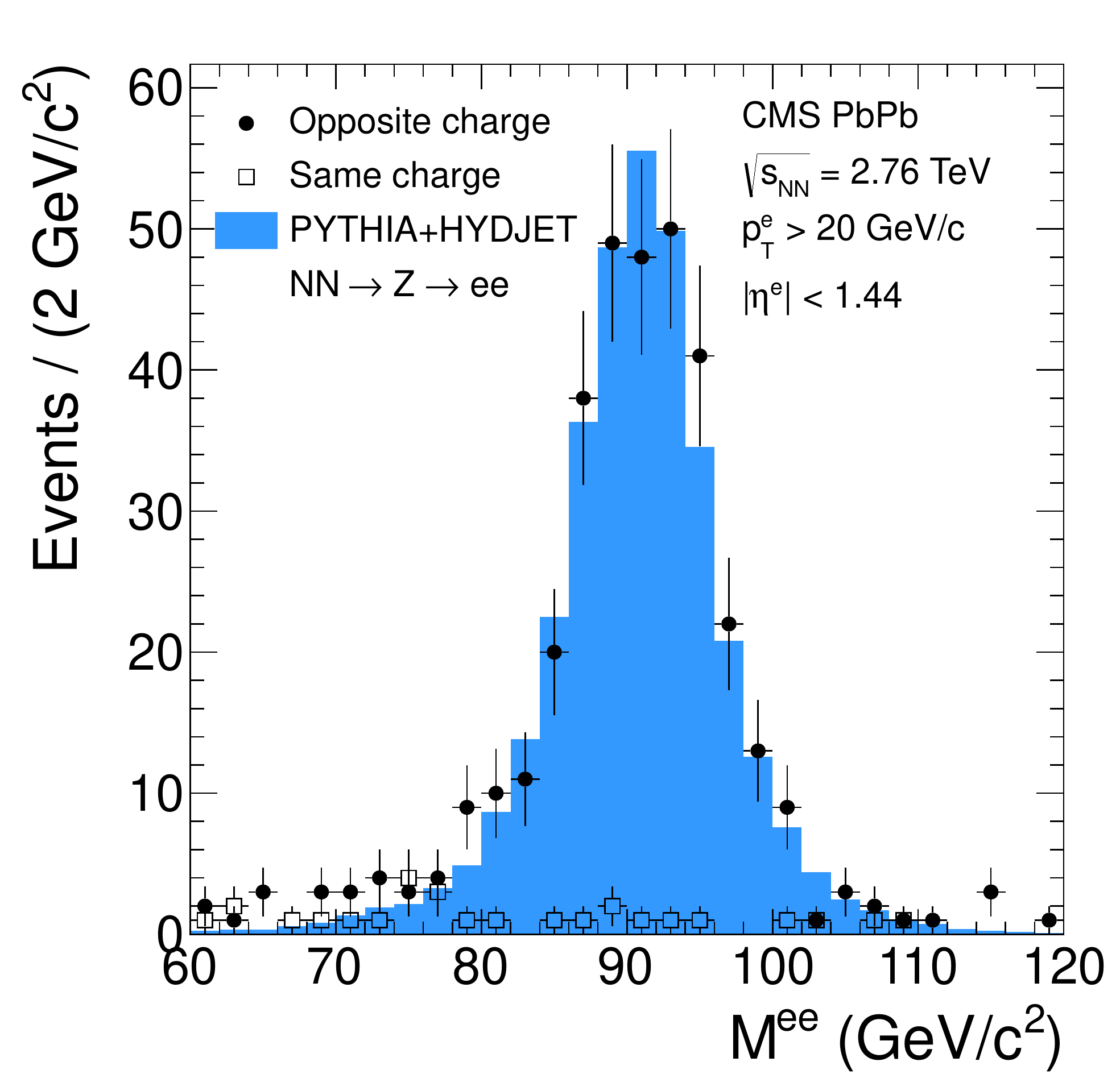}
  \includegraphics[width=0.45\textwidth]{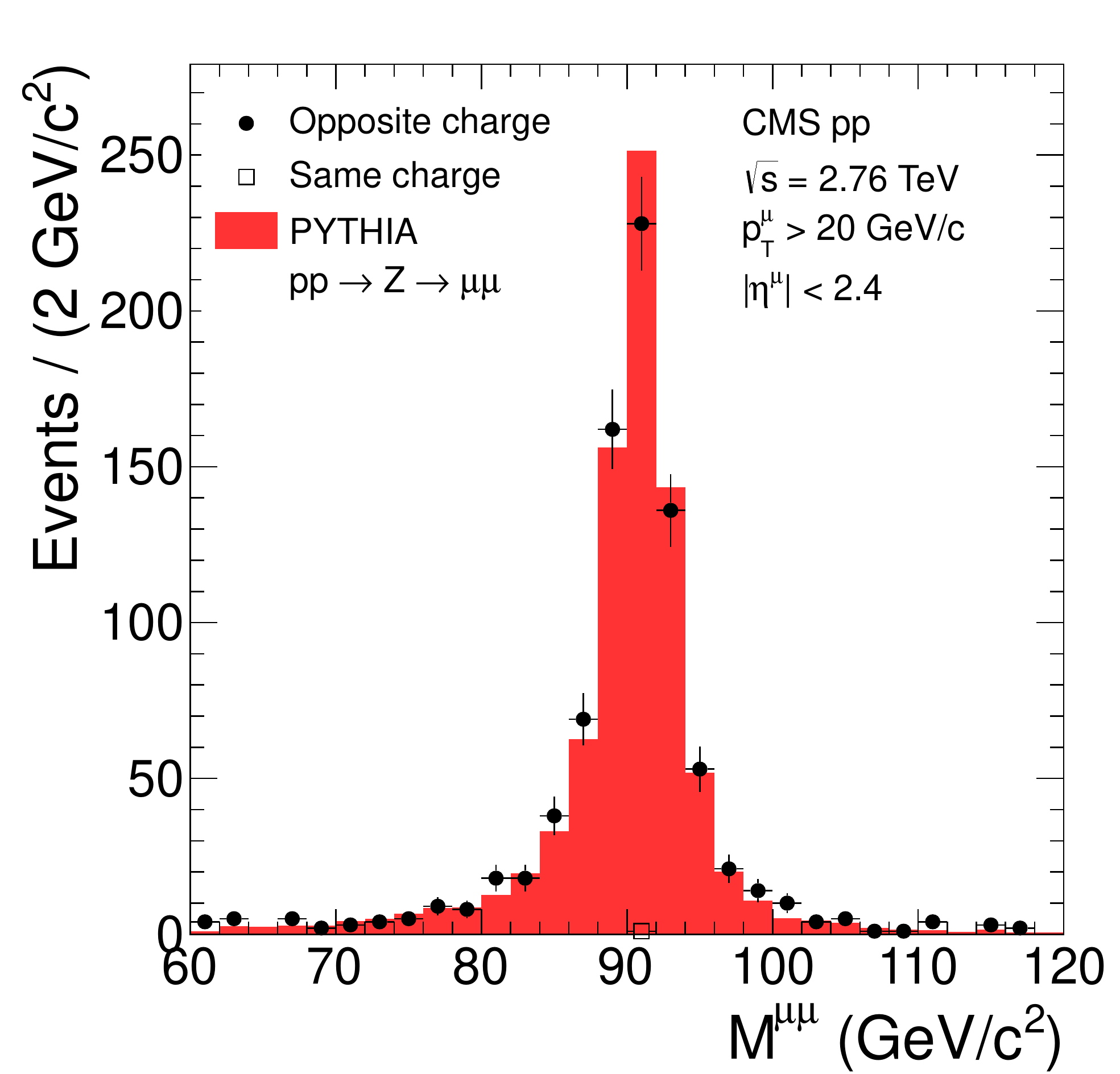}
  \includegraphics[width=0.45\textwidth]{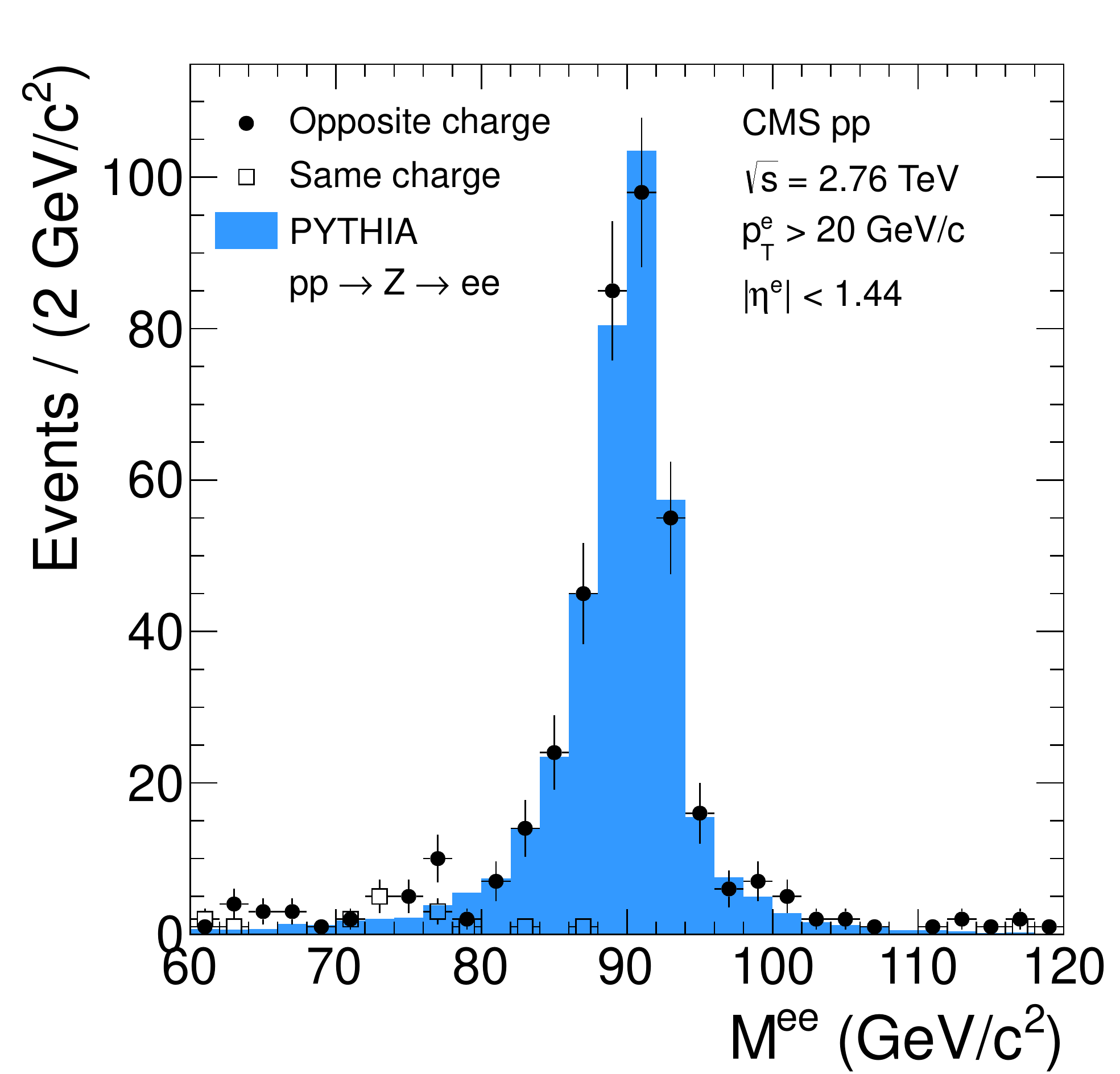}
  \caption{Dimuon invariant mass spectra for muons with $\abs{\eta^\Pgm}<2.4$ and $\pt^\mu>20$\GeVc in PbPb (top left) and in pp (bottom left) collisions and dielectron invariant mass spectra for electrons with $\abs{\eta^\Pe} < 1.44$ and $\pt^{\Pe}>20$\GeVc in PbPb (top right) and in pp (bottom right) collisions. Full black circles represent opposite-charge lepton pair events and open black squares represent same-charge lepton pair events. Superimposed and normalized to the number of Z boson candidates in data is the MC simulation from \PYTHIA $\mathrm{NN}\to\cPZ\to \Pgmp\Pgmm$ or $\Pep\Pem$, where N is a nucleon from the proper mix of protons and neutrons, embedded in \textsc{Hydjet} simulated events for the PbPb case, and $\Pp\Pp\to\cPZ\to \Pgmp\Pgmm$ or $\Pep\Pem$ for the pp case.}
  \label{fig:InvMassMuon}
  \end{center}
\end{figure}

Figure~\ref{fig:InvMassMuon} shows the dimuon and the dielectron invariant mass spectra in the 60--120\GeVcc mass range after applying acceptance and selection criteria in PbPb and pp collisions. The filled histograms are from the MC simulation described in Section~\ref{sec:acceff}. In the PbPb sample (top row), 1022 dimuon events (left column) with opposite-charge pairs (OC, black solid circles) and no events with same-charge pairs (SC, black open squares) are found in the Z boson mass range. The pp sample (bottom row) has 830 OC muon pairs and 1 SC pair. In the more restricted dielectron $y$ range (right column), 328 (388) OC pairs are found in the PbPb (pp) data sample, with 27 (17) SC pairs. The increased rate of SC pairs in the dielectron channel results from higher rates of electron misreconstruction and charge misidentification. The charge misidentification rate is estimated for electrons and results in a 1\% correction for Z bosons in PbPb collisions. The SC lepton pairs provide a measurement of the combinatorial background, which is negligible (at the 0.1\% level) in the muon channel and about 8\% (4\%) in the electron channel for PbPb (pp) data. The number of Z boson candidates is taken as the $\mathrm{OC}-\mathrm{SC}$ difference. The remaining background contamination is found to be less than 1\% and is calculated using a sideband fitting method described in Section~\ref{SystUncertainties}. These sources of background include \bbbar and \ccbar pairs, $\cPZ\to\tau^{+}\tau^{-}$, and combinations of charged leptons from W-boson decays with an additional misidentified lepton in the event.

\subsection{Acceptance and efficiency}
\label{sec:acceff}

In order to correct yields for the acceptance and efficiency in the PbPb analysis, the electroweak processes $\cPZ\to \Pgmp\Pgmm$ and $\cPZ\to \Pep\Pem$ have been simulated using the \PYTHIA~6.424~\cite{Ref:Pythia} generator, taking into account the proton and neutron content in the Pb nuclei. The detector response to each \PYTHIA signal event is simulated with \GEANTfour~\cite{Agostinelli:2002hh} and then embedded in a realistic heavy-ion background event. These background events are produced with the \textsc{Hydjet} 1.8 event generator~\cite{Lokhtin:2005px} and then simulated with \GEANTfour as well. The \textsc{Hydjet} parameters are tuned to reproduce the measured particle multiplicity for different centralities. The embedding is done at the level of detector hits, and the signal and background events share the same generated vertex location. The embedded events are then processed through the trigger emulation and the full event reconstruction chain. Finally, the generated Z boson \pt distribution is reweighted according to the \pt distribution obtained using the $\Pp\Pp\to\cPZ\to \Pgmp\Pgmm$ \POWHEG~\cite{Nason:2004rx,Frixione:2007vw,Alioli:2008gx,Alioli:2010xd} next-to-leading-order (NLO) event generator at 2.76\TeV with the CT10 PDF set~\cite{Lai:2010vv} that gives a reasonable description of the 7\TeV measurement~\cite{Chatrchyan:2011wt}. The distribution of the longitudinal position of the primary vertex is reweighted to match the one observed in collision data.

For the pp data sample, Z boson events are generated with the \PYTHIA 6.424 generator with tune Z2* that matches the charged particle multiplicity measured by CMS at $\sqrt{s}$ values of 0.9, 2.36, and 7\TeV~\cite{Chatrchyan:2013gfi}. The generated Z boson \pt distribution is reweighted according to the same \POWHEG \pt distribution used in PbPb. These generated events are reconstructed with the same software and algorithms used for the pp collision data. The longitudinal distribution of the reconstructed primary vertex matches the one in pp data.

Though computed in one step, the acceptance ($\alpha$) and efficiency ($\varepsilon$) can be split into two contributions as follows:
\begin{equation}
  \label{eq:acc}
    \alpha  = \frac{N^\cPZ  \left(\abs{y_\cPZ^{\mu\mu(\Pe\Pe)}}<2.0 (1.44),\abs{\eta^{\mu(\Pe)}} < 2.4\,(1.44), \pt^{\mu(\Pe)} \geq 20\GeVc\right) }{ N^\cPZ  \left(\abs{y_\cPZ^{\mu\mu(\Pe\Pe)}}<2.0\,(1.44) \right) },
\end{equation}
\begin{equation}
  \label{eq:eff}
     \varepsilon =   \frac{N^\cPZ \left(\abs{y_\cPZ^{\mu\mu(\Pe\Pe)}}<2.0\,(1.44),\abs{\eta^{\mu(\Pe)}} < 2.4(1.44), \pt^{\mu(\Pe)} \geq 20\GeVc\text{, quality requirements}\right)}{N^\cPZ \left(\abs{y_\cPZ^{\mu\mu(\Pe\Pe)}}<2.0\,(1.44),\abs{\eta^{\mu(\Pe)}} < 2.4(1.44), \pt^{\mu(\Pe)} \geq 20\GeVc\right)},
\end{equation}
where $N^\cPZ(\ldots)$ is the number of Z bosons satisfying the restrictions listed in the parentheses, and $y_\cPZ^{\mu\mu(\Pe\Pe)}$ is the rapidity of the dimuon (dielectron) system. The $\varepsilon$ factor reflects the reconstruction, trigger and selection efficiency of Z boson candidates. The corrections are calculated for each Z boson rapidity, \pt, or event centrality bin and the corresponding selection is applied to both the numerator and denominator.

For the Z boson \pt distributions, because of the rapidly falling \pt spectrum and the finite momentum resolution of the detector, an unfolding technique based on the inversion of a response matrix created from large simulation samples is first applied to data, similar to the one used in Ref.~\cite{Chatrchyan:2011wt}, before applying the acceptance and efficiency correction based on the generated quantities. The \pt resolution of the Z in the dimuon (dielectron) channel varies from 7\% (22\%) at low Z \pt to 2.5\% (2.5\%) at high Z \pt. Due to the correlations between neighboring bins, the variance in the statistical uncertainties increases, which is taken into account in the quoted statistical uncertainties. Using unfolding in rapidity is not needed as the shape of the $y$ spectrum is relatively flat.

For the Z boson rapidity distributions, the efficiency corrections are done such that the denominator is the number of generated Z boson events that survive the selection on kinematic quantities and binned in the kinematic quantities of the generated Z boson. The numerator is the number of reconstructed dimuons (dielectrons) after applying the selection criteria to the dimuon (dielectron) reconstructed quantities and binning based on those reconstructed quantities. This choice folds the minimal resolution effects in $y$ into the efficiency correction.

The overall acceptance is approximately 70 (50)\% in the muon (electron) rapidity ranges, and the overall detection efficiency is approximately 85\,(55)\% in PbPb and 90\,(80)\% in pp collisions.

\subsection{Systematic uncertainties}
\label{SystUncertainties}

The total systematic uncertainty in the Z boson yield in PbPb collisions is estimated by adding in quadrature the different contributions. The uncertainty on the combined trigger, reconstruction and selection efficiency is 1.8\%\,(7.4\%) for the dimuon (dielectron) channel. This estimate is based on the \emph{tag-and-probe} technique for measuring single-particle reconstruction, identification, and trigger efficiencies, which is done in a way similar to the method described in Ref.~\cite{Chatrchyan:2012np} and is dominated by the statistical uncertainty in the data.

The uncertainties coming from the acceptance corrections are less than 2\%, as estimated by applying to the generated Z boson \pt and $y$ distributions a weight that varies linearly between 0.7 and 1.3 over the ranges $\pt<100$\GeVc and $\abs{y}<2.0\,(1.44)$ for the dimuon (dielectron) channel. The \pt-dependent uncertainty arising from the resolution unfolding is less than $1\%$, as estimated by varying the generated Z boson \pt distribution using the same method. The energy scale of the electrons and muons relies heavily on the information from the track (in combination with the calorimeter and muon chambers), which decreases the uncertainty of the energy scale. The energy scale uncertainty is less than 1\% for the final $R_\mathrm{AA}$.

The systematic uncertainty from the remaining backgrounds from other physical sources, such as heavy-flavour semi-leptonic decays, is estimated by fitting the lower dilepton mass range for the data (with the Drell--Yan contribution subtracted) with an exponential function and extrapolating the fit to higher masses. This fit gives a conservative systematic uncertainty of 0.5 (2.0)\% in the dimuon (dielectron) channel.

In pp collisions, the largest systematic uncertainty in the differential cross sections comes from the luminosity determination, which is 3.7\%. The other sources of systematic uncertainties are similar to the ones described for PbPb collisions.

For the $R_\mathrm{AA}$ measurement described in Section~\ref{sec:R_AA}, the Z boson cross section in pp collisions is scaled by $T_\mathrm{AA}$ in order to compare with the corrected yields in PbPb collisions. The uncertainties in $T_\mathrm{AA}$ are derived by varying the Glauber model parameters and the minimum bias event selection efficiency within their uncertainties, resulting in 6.2\% relative uncertainty for centrality-integrated quantities.

The systematic uncertainties for the $\cPZ\to \Pgmp\Pgmm$ and $\Pep\Pem$ channels for both yields in PbPb and pp collisions are summarized in Table~\ref{tab:systematics}.

\begin{table}[h!]
\centering
\topcaption{Summary of systematic uncertainties in the $\cPZ\to \Pgmp\Pgmm $ and $\Pep\Pem$ yields. PbPb values correspond to the full 0--100\% centrality range. $N_\mathrm{MB}$ is the number of MB events corrected for the trigger efficiency.}
\label{tab:systematics}
\begin{tabular}{c|c|c|c|c}

 \multicolumn{1}{c}{}&  \multicolumn{2}{c|}{$\cPZ\to\Pgmp\Pgmm$ } & \multicolumn{2}{c}{ $\cPZ\to \Pep\Pem$} \\ \cline{2-5}
  Source & PbPb & pp & PbPb & pp \\ \hline
  Combined efficiency & 1.8\% & 1.9\% & 7.4\% & 7.7\% \\
  Acceptance & 0.7\% & 0.7\% & 0.7\% & 0.7\% \\
  Background & 0.5\% & 0.1\% & 2.0\% & 1.0\% \\
  $N_\mathrm{MB}$ & 3.0\% & \textemdash & 3.0\% & \textemdash \\ \hline
  $T_\mathrm{AA}$  ($N_\mathrm{MB}$ included) & 6.2\% & \textemdash  & 6.2\% & \textemdash  \\
  Integrated luminosity ($L_\text{int}$) & \textemdash & 3.7\%  & \textemdash & 3.7\% \\ \hline
  Overall (without $T_\mathrm{AA}$ or $L_\text{int}$) & 3.6\% & 2.0\% & 8.3\% & 7.8\% \\ \hline
  Overall & 6.5\% & 4.2\% & 9.9\% & 8.6\% \\ \hline
\end{tabular}
\end{table}

\section{Results}
\subsection{Z boson production cross section in pp collisions}
\label{sec:ZYieldspp}

\begin{figure}[ht]
 \begin{center}
   \includegraphics[width=0.45\textwidth]{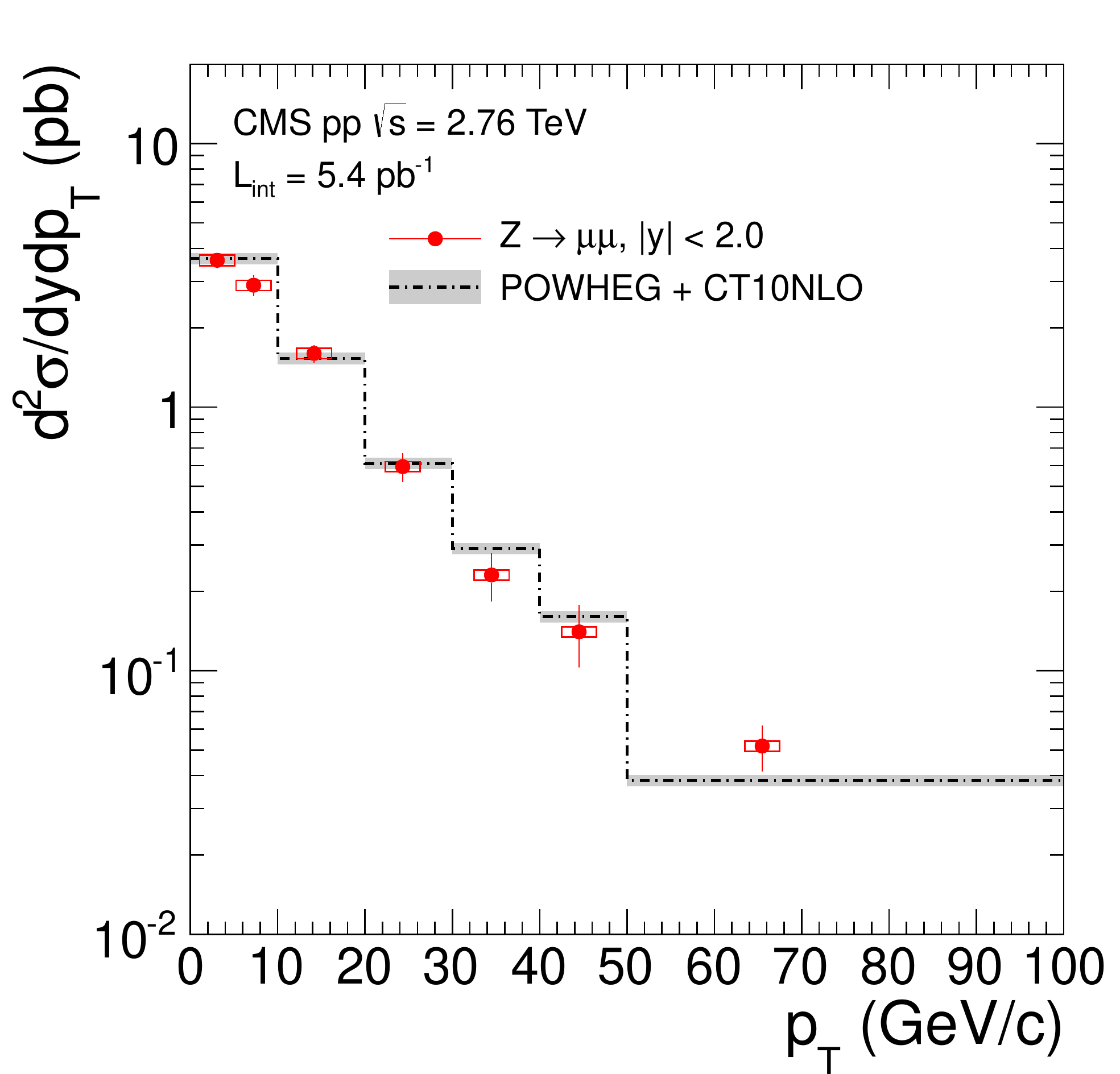}
   \includegraphics[width=0.45\textwidth]{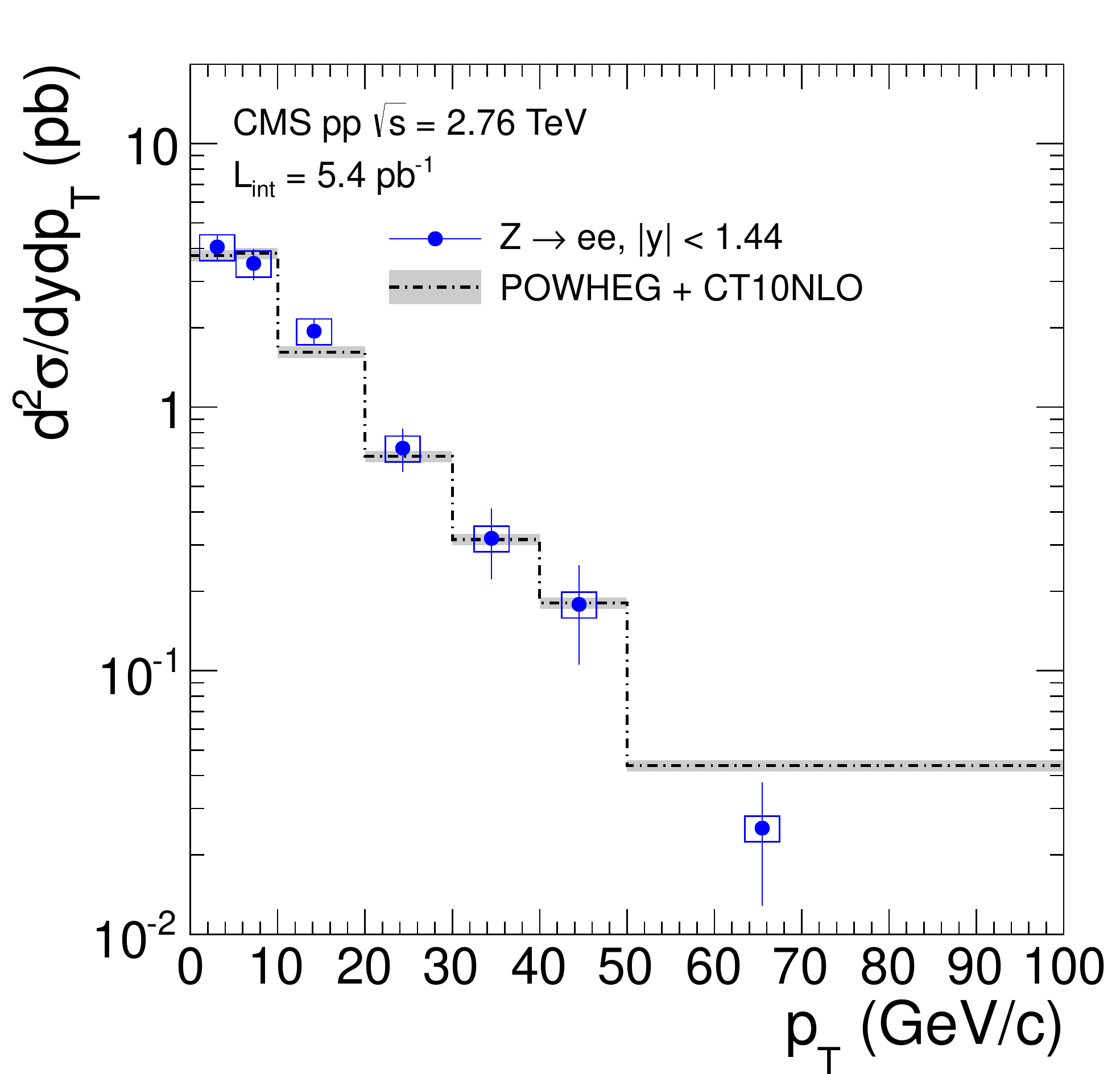}
   \includegraphics[width=0.45\textwidth]{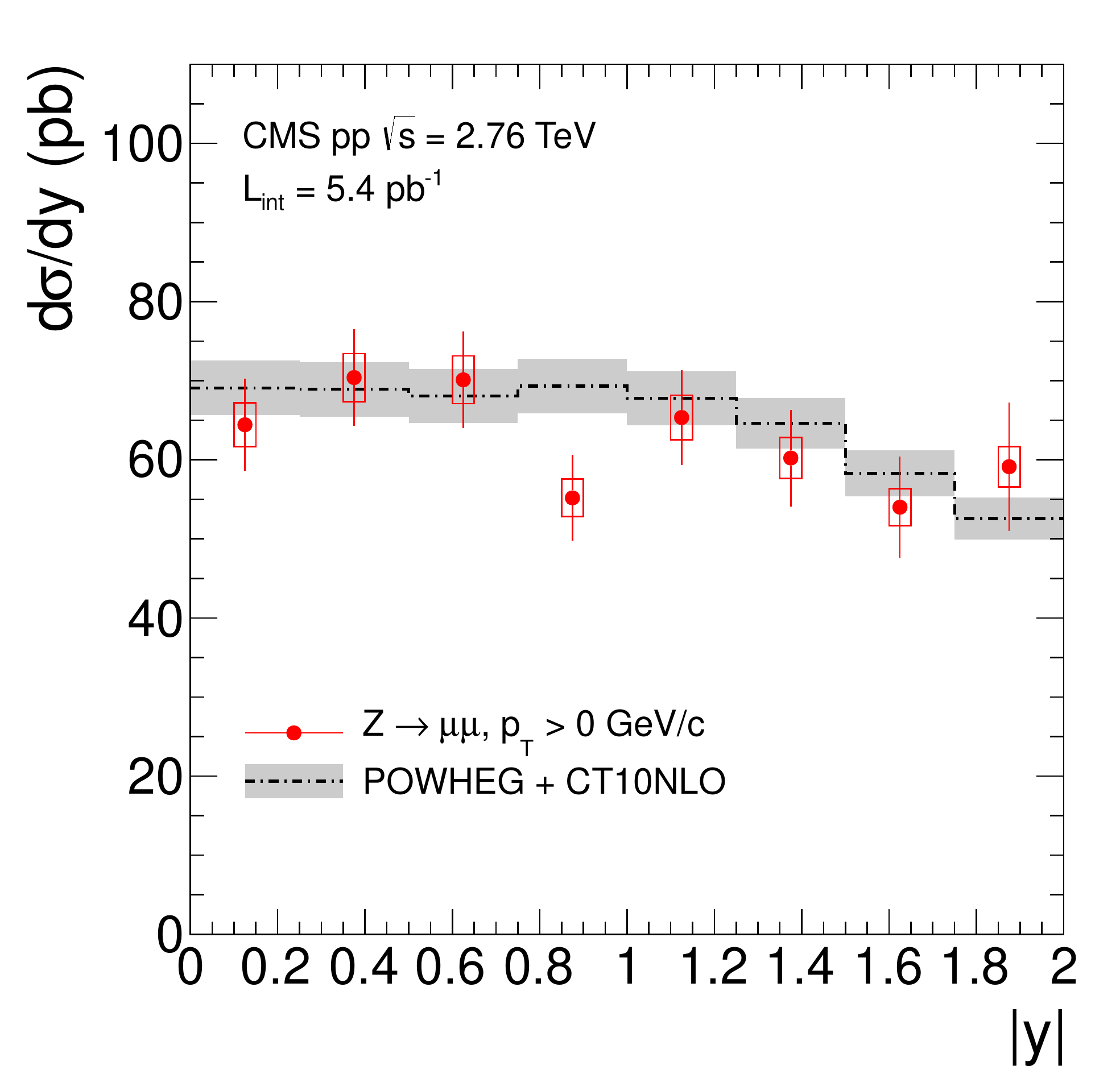}
   \includegraphics[width=0.45\textwidth]{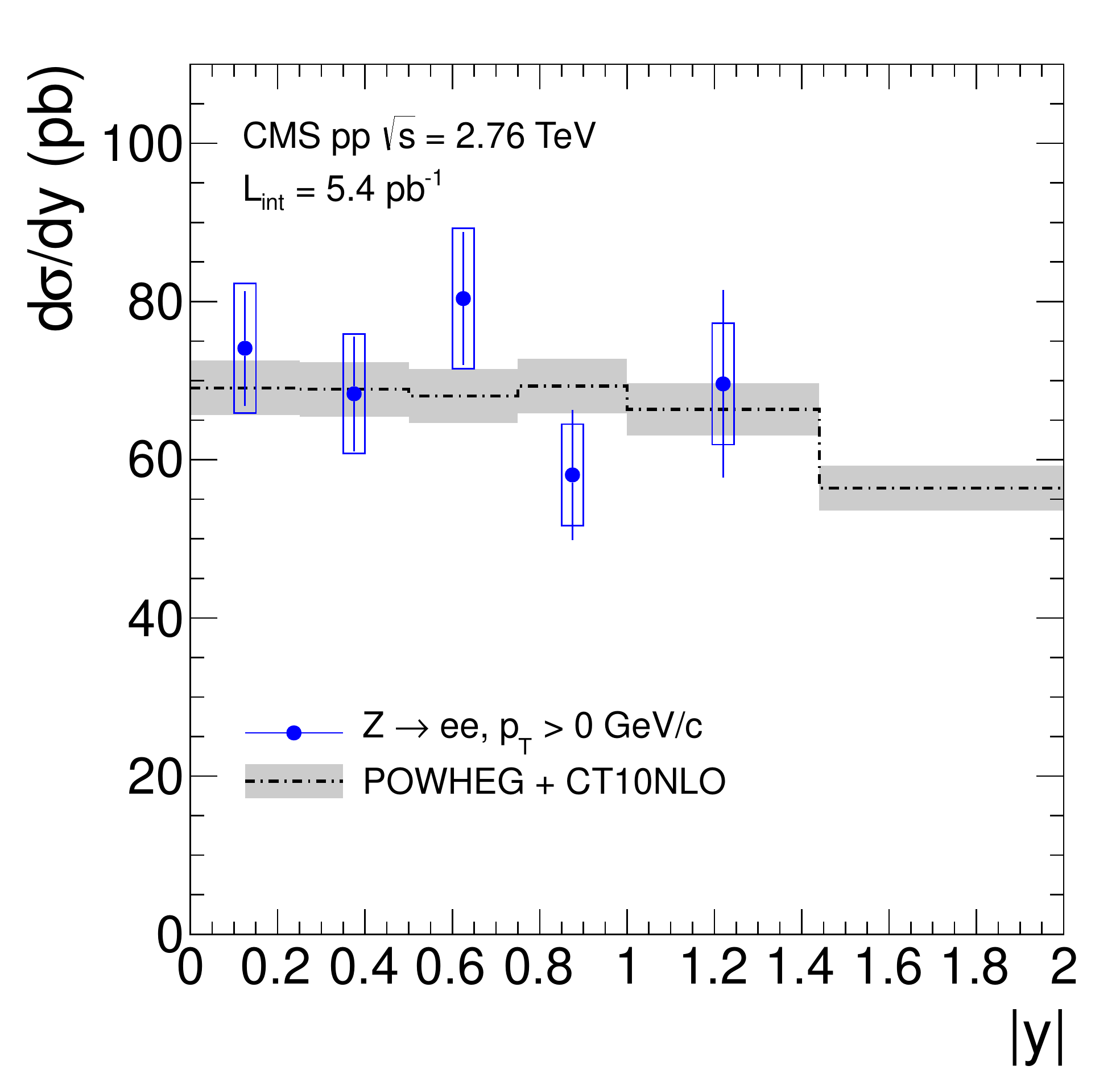}
   \caption{The measured Z boson production cross section in pp collisions as a function of the Z boson \pt (top) and $y$ (bottom) for the dimuon (left) and the dielectron (right) decay channels. Results are compared with $\Pp\Pp\to\cPZ\to \ell^{+}\ell^{-}$ \POWHEG predictions. Vertical lines (boxes) correspond to statistical (systematic) uncertainties. The theoretical uncertainty of 5\% assumed for the \POWHEG reference curve is shown by the grey band.}
    \label{fig:resultsppxsectionpT_y}
 \end{center}
\end{figure}

The differential $\Pp\Pp\to\cPZ\to \Pgmp\Pgmm$ and $\Pep\Pem$ cross sections as a function of \pt and $y$ of the Z boson candidates, selected in the mass range between 60--120\GeVcc mass and within $\abs{y}<2.0\,(1.44)$ in the dimuon (dielecton) channel, are obtained from the pp collision data at $\sqrt{s} = 2.76$\TeV. These distributions are shown in Fig.~\ref{fig:resultsppxsectionpT_y}. For pp collisions, the cross section is calculated by dividing the corrected yields by the calibrated integrated luminosity. Overall, the differential cross sections agree with the \POWHEG theoretical predictions. Higher-order corrections to the cross sections predicted by this generator amount to 3\%~\cite{Catani:2009sm}. Typical next-to-next-to-leading-order calculations also have a 3\% uncertainty in the proton PDFs and are found to agree with 7~and~8\TeV pp data, as reported in Refs.~\cite{CMS:2011aa, Chatrchyan:2014mua}. Therefore, the \POWHEG reference has a typical uncertainty of 5\% as indicated by the grey band.

\subsection{Z boson yields in PbPb collisions vs. \texorpdfstring{\pt, $y$,}{pt, y,} and centrality}
\label{sec:ZyieldsPbPb}

The yield of Z bosons has been measured in PbPb collisions as a function of event centrality, Z boson $y$ and \pt, and then compared to that in pp collisions simulated using \POWHEG, scaled by an average nuclear overlap function ($T_\mathrm{AA}$), as described below and discussed in Ref.~\cite{Miller:2007ri}.
Simulated pp collisions from \POWHEG are used in these first comparisons because of their higher statistical precision. As shown in Section~\ref{sec:ZYieldspp}, the pp data are consistent with the simulations. A direct comparison of PbPb and pp data is shown in Section~\ref{sec:R_AA}.

\begin{figure}[ht]
  \begin{center}
    \includegraphics[width=0.45\textwidth]{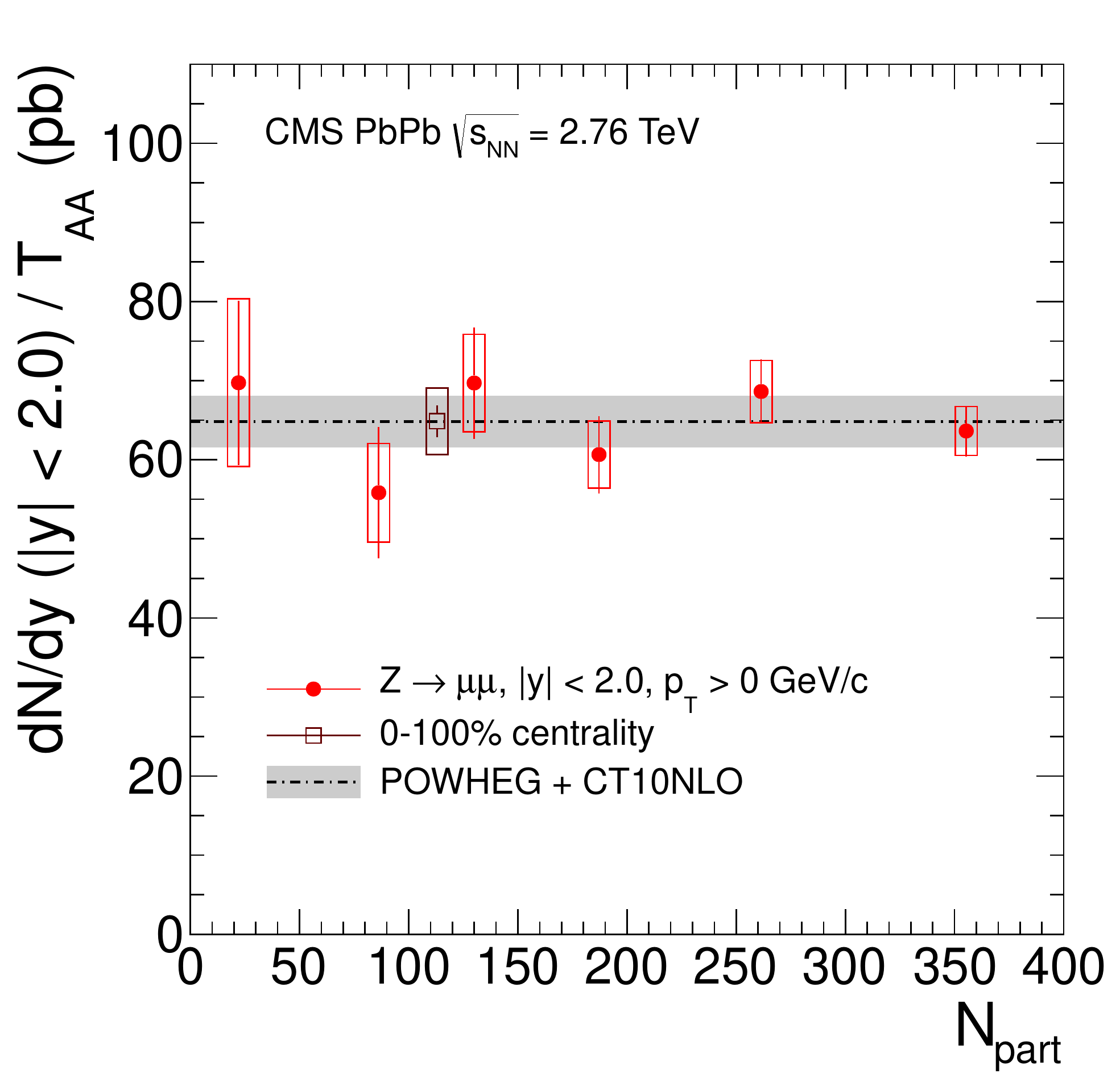}
     \includegraphics[width=0.45\textwidth]{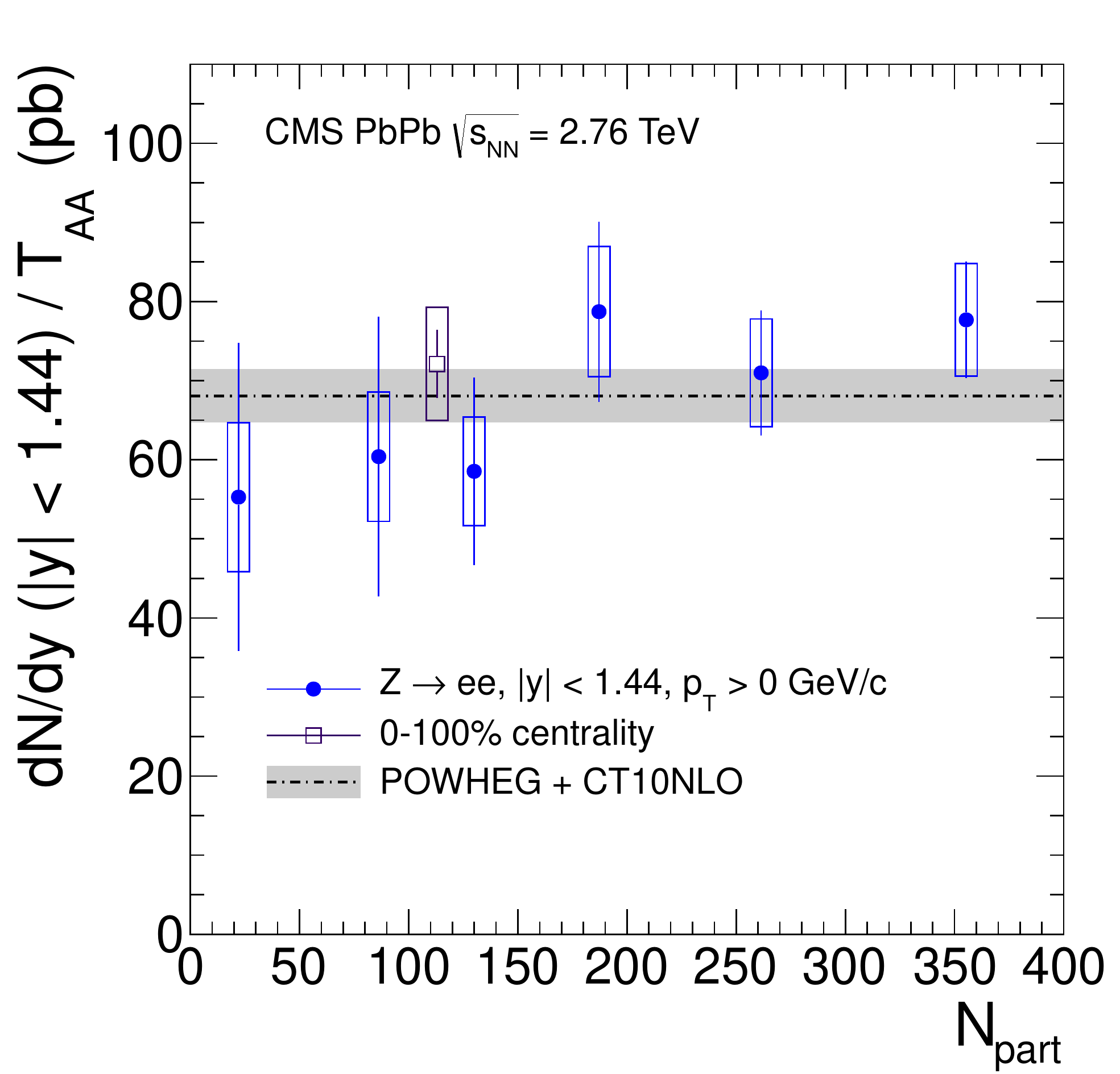}
      \caption{Event centrality dependence of the $\cPZ\to\Pgmp\Pgmm$ (left) and $\cPZ\to\Pep\Pem$ (right) yields per MB event in PbPb collisions, divided by the expected average nuclear overlap function, $T_\mathrm{AA}$, which is directly comparable to the $\Pp\Pp\to\cPZ\to \ell^{+}\ell^{-}$ cross section predicted by the \POWHEG generator displayed as a black dash-dotted line. On the horizontal axis, event centrality is depicted as the average number of participating nucleons, $N_\text{part}$ (see Table~\ref{tab:glauber}). Vertical lines (boxes) correspond to statistical (systematic) uncertainties. The theoretical uncertainty of 5\% assumed for the \POWHEG reference curve is shown by the grey band.}
    \label{fig:yieldvsNpart}
  \end{center}
\end{figure}

\begin{figure}[ht]
  \begin{center}
    \includegraphics[width=0.45\textwidth]{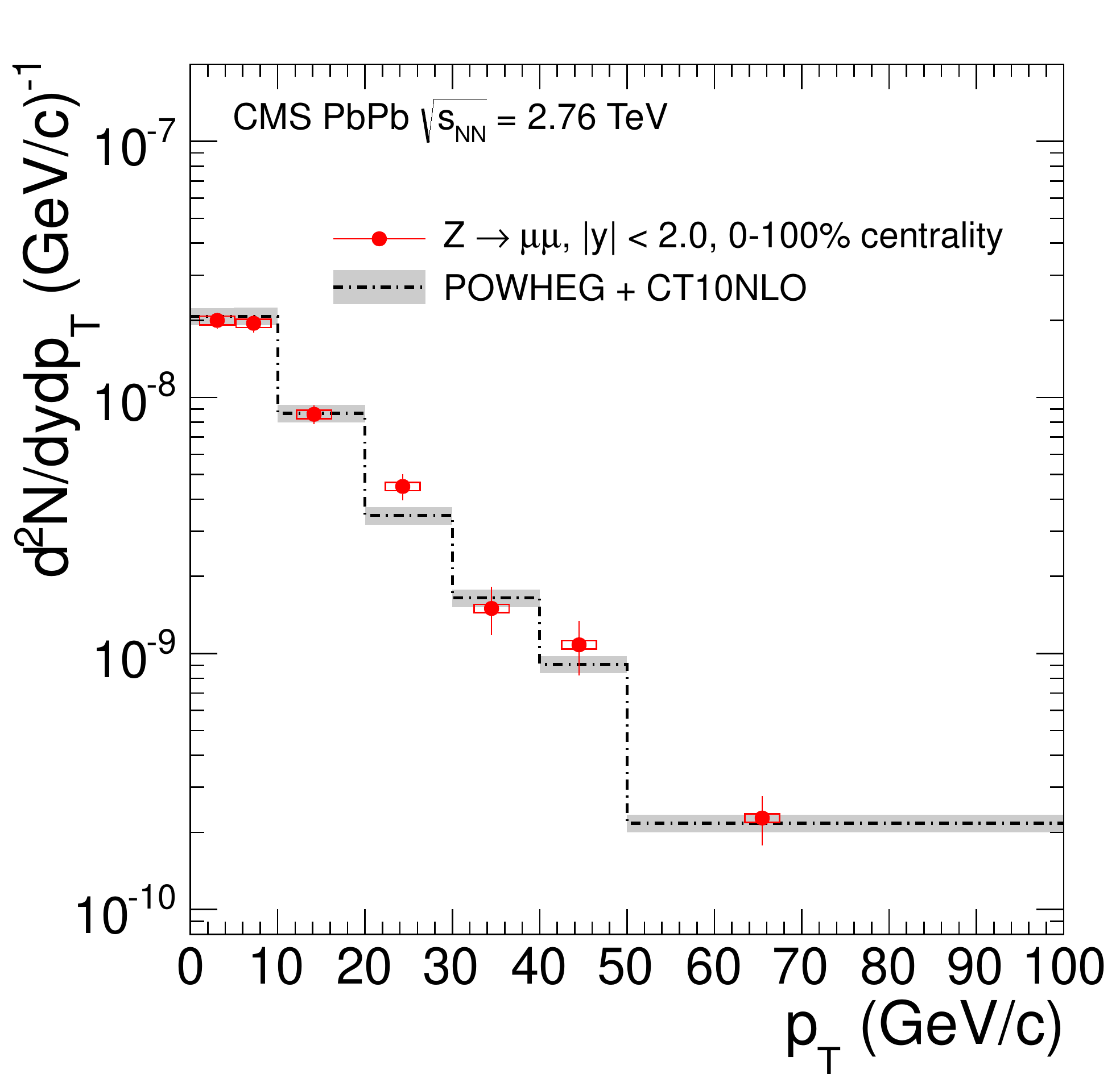}
    \includegraphics[width=0.45\textwidth]{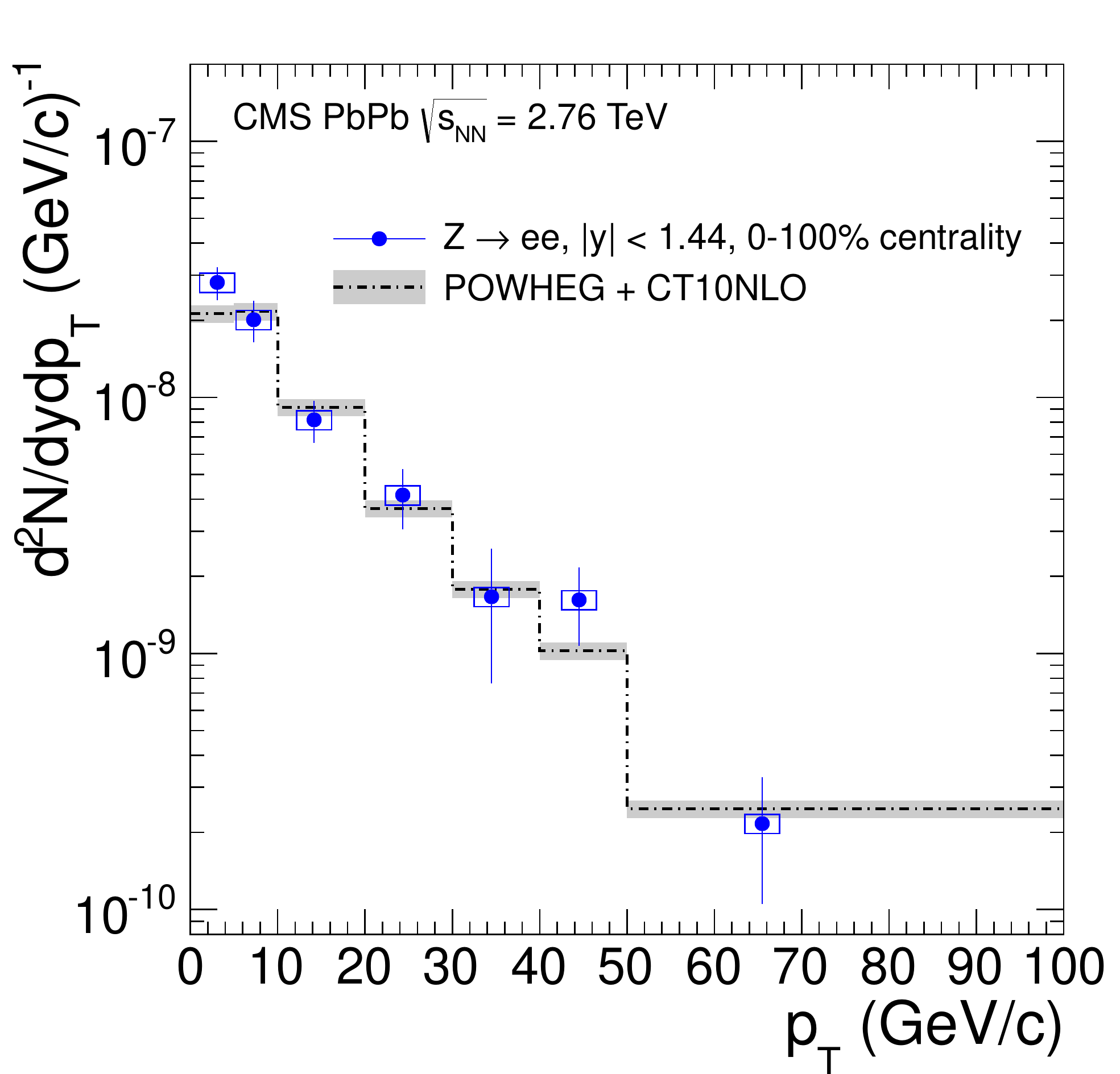}
    \includegraphics[width=0.45\textwidth]{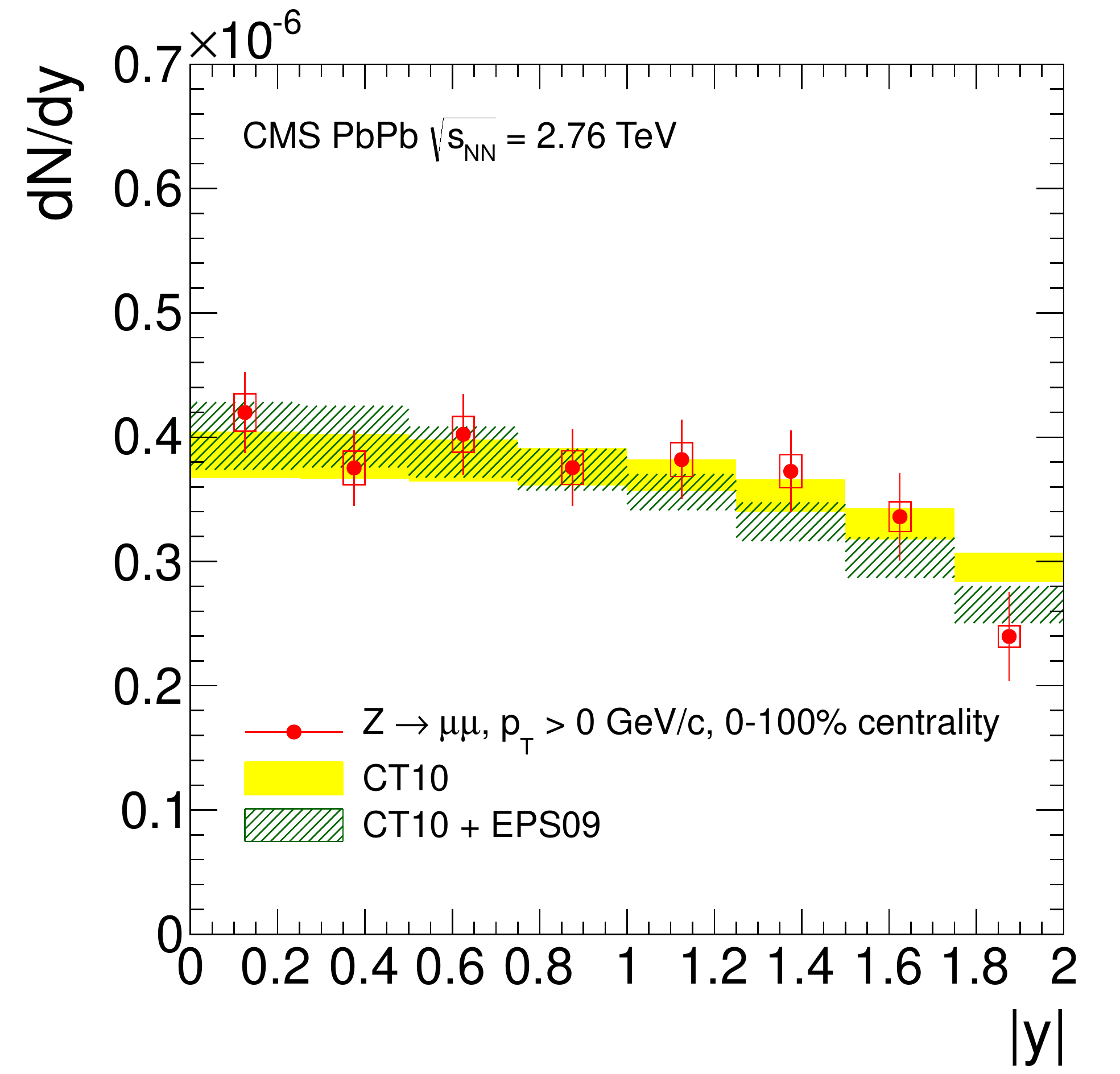}
    \includegraphics[width=0.45\textwidth]{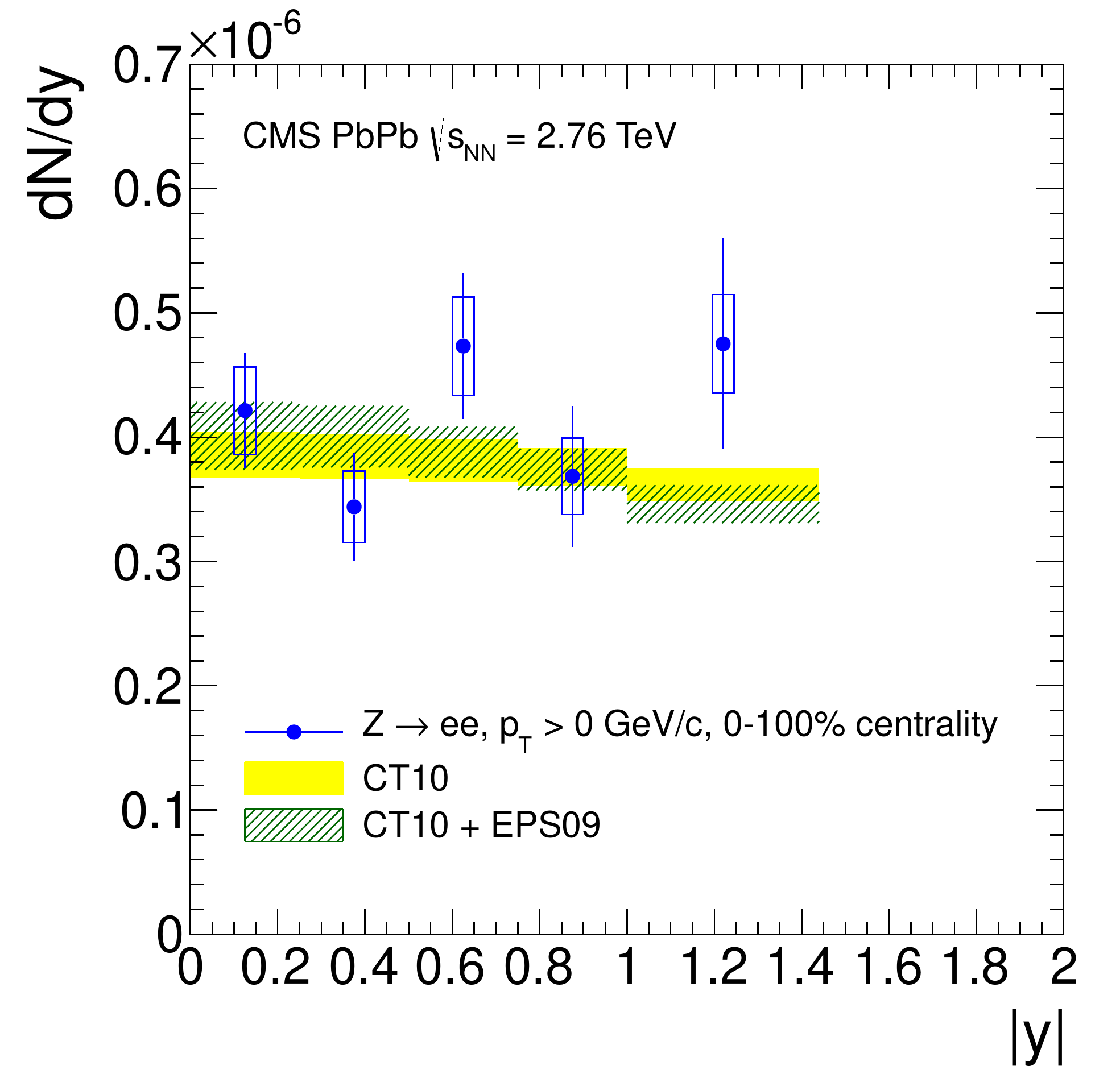}
    \caption{The measured Z boson yields per MB event in PbPb collisions as a function of the Z boson \pt (top) and $y$ (bottom) for the dimuon (left) and the dielectron (right) decay channels. The yields are compared with $\Pp\Pp\to\cPZ\to \ell^+\ell^-$ \POWHEG predictions scaled by the 0--100\% centrality averaged $T_\mathrm{AA}$. The light gray bands in the results vs. \pt represent the theoretical uncertainty of 5\% assumed for the \POWHEG reference curve together with the uncertainty of 6.2\% due to the $T_\mathrm{AA}$ scaling. The results vs. $y$ are compared to predictions with (green dark band) and without (yellow light band) nuclear modification effects. Vertical lines (boxes) correspond to statistical (systematic) uncertainties.}
  \label{fig:yieldvspT_y}
  \end{center}
\end{figure}

The data are divided into independent ranges: 6 in event centrality, 8\,(5) in $y$ for the dimuon (dielectron) channel, and 7 in the dilepton \pt. The results are presented in Figs.~\ref{fig:yieldvsNpart} and ~\ref{fig:yieldvspT_y}. The yields of $\cPZ\to \ell^{+}\ell^{-}$ (where $\ll$ is $\mu$ or \Pe) per MB event, per unit of $y$ ($\rd{}N^\cPZ_\mathrm{PbPb}/\rd{}y$), and per \pt bin ($\rd^2N^\cPZ_\mathrm{PbPb}/\rd{}y\,\rd\pt$) are computed using the following equations:
\begin{linenomath}
\begin{equation}
\label{eq:yields}
 \frac{\rd{}N^\cPZ_\mathrm{PbPb}}{\rd{}y} = \frac{N_\mathrm{PbPb}(\cPZ \to \ell^+ \ell^-)}{\alpha \varepsilon N_\mathrm{MB} \Delta y} \text{  or  }
 \frac{\rd^2N^\cPZ_\mathrm{PbPb}}{\rd{}y\,\rd\pt} = \frac{N_\mathrm{PbPb}(\cPZ \to \ell^+ \ell^-)}{\alpha \varepsilon N_\mathrm{MB} \Delta y \Delta \pt}.
\end{equation}
\end{linenomath}

Here $N_\mathrm{PbPb}(\cPZ \to \ell^+ \ell^-)$ is the number of Z boson candidates, divided into bins of \pt, $y$, and centrality, found in the dimuon or dielectron invariant mass range of 60--120\GeVcc; $N_\mathrm{MB}$ is the number of corresponding MB events corrected for the trigger efficiency, namely $(1.16\pm 0.03)\times10^9$ events; $\alpha$ and $\varepsilon$ are acceptance and efficiency corrections (see Section~\ref{sec:acceff}); $\Delta y$ and $\Delta \pt$ are the two bin widths under consideration. When the Z boson yield is divided into centrality bins, $N_\mathrm{MB}$ is multiplied by the corresponding fraction of the MB cross section included in the bin.

Fig.~\ref{fig:yieldvsNpart} shows the centrality dependence of the Z boson production in PbPb collisions. The $\rd{}N^\cPZ_\mathrm{PbPb} / \rd{}y$ yields per MB event are divided by the nuclear overlap function $T_\mathrm{AA}$. This quantity is proportional to the number of inelastic nucleon-nucleon collisions $N_\text{coll} = T_\mathrm{AA} \times \sigma_\mathrm{NN}^\text{inel}$, where $\sigma_\mathrm{NN}^\text{inel}=64\pm5$\unit{mb} is the inelastic nucleon-nucleon cross section. The $T_\mathrm{AA}$ uncertainties are included in the systematic uncertainties depicted as boxes around the data in Fig.~\ref{fig:yieldvsNpart}. On the horizontal axis, the event centrality is translated to the average number of participants ($N_\text{part}$) as shown in Table~\ref{tab:glauber}, using the same Glauber model.

No strong centrality dependence is observed for the yield $\rd{}N^\cPZ_\mathrm{PbPb} / \rd{}y \times 1/T_\mathrm{AA}$. The centrality-integrated value is displayed as an open square for each channel in Fig.~\ref{fig:yieldvsNpart}. For comparison, the dash-dotted line on the plots shows the cross section of the $\Pp\Pp\to\cPZ\to \ell^{+}\ell^{-}$ process provided by the \POWHEG generator interfaced with the \PYTHIA~6.424 parton-shower generator.

For the \pt dependence and $y$ dependence of the Z boson yields, the data are integrated over centrality; therefore the \POWHEG reference is multiplied by the 0-100\% centrality averaged $T_\mathrm{AA} = 5.67 \pm 0.32\mbinv$, as provided by the Glauber model described above. By construction, this centrality-averaged $T_\mathrm{AA}$ is equal to $A^{2}$/$\sigma_\mathrm{PbPb}^\text{inel}$, where $A = 208$ is the Pb atomic number and $\sigma_\mathrm{PbPb}^\text{inel} = 7.65 \pm 0.42$\unit{barns} is the total PbPb inelastic cross section computed from the same Glauber model.

Fig.~\ref{fig:yieldvspT_y} shows the distribution $\rd^2N^\cPZ_\mathrm{PbPb} /\rd{}y\,\rd\pt$ as a function of the dilepton \pt and the invariant yield as a function of rapidity, $\rd{}N^\cPZ_\mathrm{PbPb} /\rd{}y$ compared to theoretical predictions, as follows. The results vs. \pt are compared to \POWHEG, while the results vs. $y$ are compared to predictions from Paukkunen and Salgado~\cite{Isospin} which do not incorporate nuclear PDF modifications to the unbound proton/nucleon PDFs (yellow light band) and those that do (green dark band) through the nuclear PDF set EPS09~\cite{Eskola:2009uj}. No strong deviations from these absolutely-normalized references are observed.

The Z boson yields in PbPb collisions have been compared with various theoretical predictions, including PDFs that incorporate nuclear effects. The calculated yields are found to be consistent with the results. Therefore, we deduce that Z boson production scales with the number of inelastic nucleon-nucleon collisions. Furthermore, nuclear effects such as isospin or shadowing are small compared to the statistical uncertainties, hence it is not possible to discriminate among these nuclear effects with the available data.

\subsection{Nuclear modification factor}
\label{sec:R_AA}

\begin{figure}[ht]
  \begin{center}
    \includegraphics[width=0.45\textwidth]{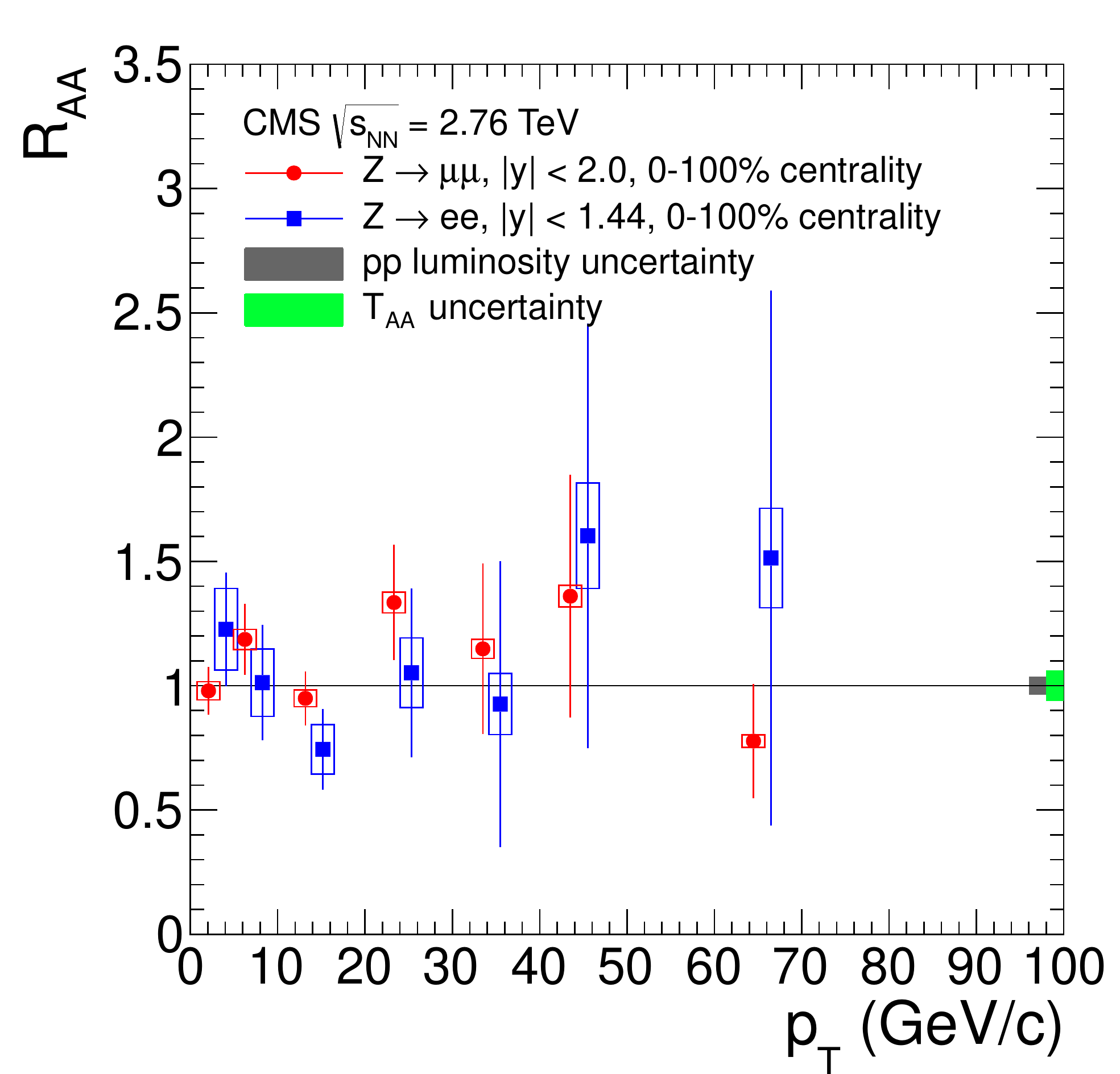}
    \includegraphics[width=0.45\textwidth]{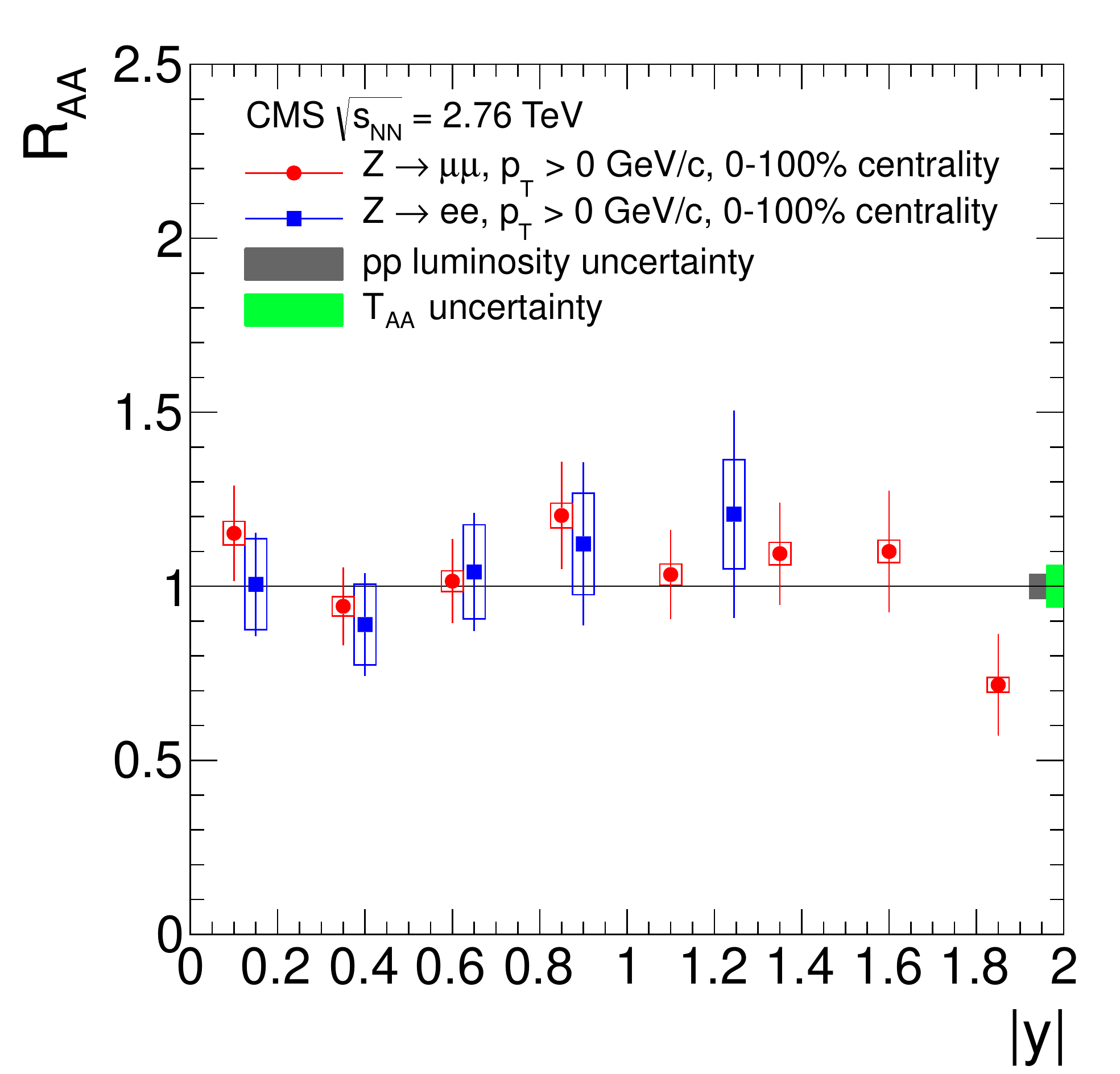}
    \includegraphics[width=0.45\textwidth]{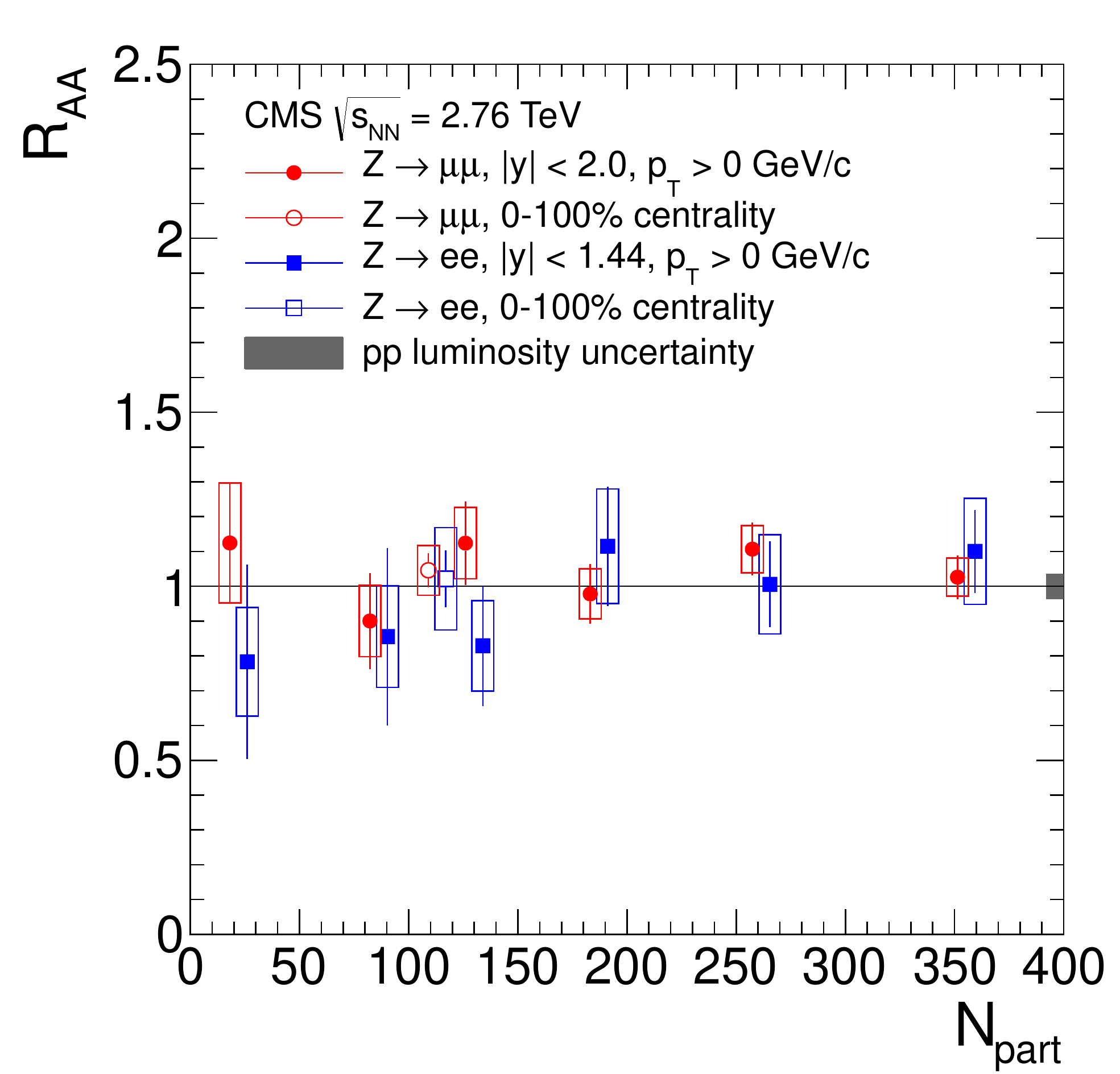}
 \caption{The $R_\mathrm{AA}$ distribution for the $\cPZ\to \Pep\Pem$ (blue squares) and $\cPZ\to\Pgmp\Pgmm$ (red circles) events as a function of the Z boson \pt (left), $y$ (right), and $N_\text{part}$ (bottom). For $N_\text{part}$, open points at $N_\text{part} \sim 110$ represent the centrality-integrated $R_\mathrm{AA}$. Points are shifted along the horizontal axis for clarity. The horizontal line at $R_\mathrm{AA}$ = 1 is drawn as a reference. Vertical lines (boxes) correspond to statistical (systematic) uncertainties. The grey bar at $R_\mathrm{AA} = 1$ corresponds to uncertainty in pp luminosity and the green bar corresponds to uncertainty in $T_{AA}$.}
  \label{fig:RAAvspT_y_Npart}
  \end{center}
\end{figure}

Based on PbPb and pp data at the same centre-of-mass energy, the nuclear modification factor, $R_\mathrm{AA}$, is computed for both the dimuon and dielectron channels as a function of the  Z boson \pt, $y$, and event centrality, as follows:
\begin{equation}
  \label{eq:raa}
R_\mathrm{AA} = \frac{N^\cPZ_\mathrm{PbPb}}{T_\mathrm{AA} \times \sigma^\cPZ_{\Pp\Pp}} \equiv \frac{N^\cPZ_\mathrm{PbPb}}{N_\text{coll} \times N^\cPZ_{\Pp\Pp}}
\end{equation}
where $N^\cPZ_\mathrm{PbPb}$ ($N^\cPZ_{\Pp\Pp}$) are the yields per MB event measured in PbPb (pp) collisions corrected for acceptance and efficiency, $\sigma^\cPZ_{\Pp\Pp}$ refers to the differential cross sections measured from pp collisions, $N_\text{coll}$ refers to the average number of inelastic nucleon-nucleon collisions for the appropriate centrality selection, and $T_\mathrm{AA}$ refers to the values of the nuclear overlap function as described in Section~\ref{sec:selection}. The $R_\mathrm{AA}$ values as a function of $y$, \pt and centrality are shown in Fig.~\ref{fig:RAAvspT_y_Npart} (where the points are slightly shifted along the horizontal axis for clarity: left for muons and right for electrons), and in Table~\ref{tab:all_together}.

The information in Fig.~\ref{fig:RAAvspT_y_Npart} is similar to that shown in Figs.~\ref{fig:yieldvsNpart} and~\ref{fig:yieldvspT_y} but here using pp data for comparison instead of \POWHEG simulations. The $R_\mathrm{AA}$ values show no dependence, and hence no variation in nuclear effects, as a function of \pt, $y$, or centrality in both the muon and electron channels in the kinematic range studied and within the current uncertainties.

\begin{table}[h!]
\centering
\topcaption{Nuclear modification factor ($R_\mathrm{AA}$) for the  $\cPZ\to\ell^+\ell^-$ process as a function of rapidity, \pt, and event centrality. The rapidity integrated values are shown for $\abs{y}<2.0$ for the muon channel and for $\abs{y}<1.44$ in case of the electron channel and for the combined channel. The first uncertainty is statistical and the second one is systematic.}
\label{tab:all_together}
\begin{tabular}{c|ccc}

 \multicolumn{4}{c}{$R_\mathrm{AA} $} \\ \hline
$\abs{y}$         &  Z $\to \mu^+\mu^- $ &  Z $\to \Pep\Pem $ & $\cPZ \to \ell^+\ell^- $ \\ \hline
$[0.00,0.25]$ & $ 1.17\pm 0.14\pm 0.09$ & $ 1.01\pm 0.15\pm 0.15$ & $ 1.13\pm 0.11\pm 0.09$ \\
$[0.25,0.50]$ & $ 0.96\pm 0.11\pm 0.07$ & $ 0.89\pm 0.15\pm 0.13$ & $ 0.96\pm 0.09\pm 0.08$ \\
$[0.50,0.75]$ & $ 1.03\pm 0.12\pm 0.08$ & $ 1.04\pm 0.17\pm 0.15$ & $ 1.04\pm 0.10\pm 0.09$ \\
$[0.75,1.00]$ & $ 1.22\pm 0.16\pm 0.10$ & $ 1.12\pm 0.23\pm 0.17$ & $ 1.22\pm 0.13\pm 0.10$ \\
$[1.00,1.25]$ & $ 1.05\pm 0.13\pm 0.08$ &\NA&\NA\\
$[1.25,1.50]$ & $ 1.11\pm 0.15\pm 0.09$ & \NA & \NA   \\
$[1.00,1.44]$ & \NA & $ 1.21\pm 0.30\pm 0.18$ & $ 1.14\pm 0.10\pm 0.09$ \\
$[1.50,1.75]$ & $ 1.12\pm 0.18\pm 0.09$ & \NA & $ 1.12\pm 0.18\pm 0.09$   \\
$[1.75,2.00]$ & $ 0.73\pm 0.15\pm 0.06$ & \NA & $ 0.73\pm 0.15\pm 0.06$   \\
\hline
$\pt (\GeVcns)$ & $ $ & $ $ & $ $ \\ \hline
$[0,5]$ & $ 0.99\pm 0.09\pm 0.08$ & $ 1.23\pm 0.23\pm 0.19$ & $ 0.99\pm 0.09\pm 0.08$ \\
$[5,10]$ & $ 1.20\pm 0.13\pm 0.10$ & $ 1.01\pm 0.23\pm 0.15$ & $ 1.29\pm 0.14\pm 0.11$ \\
$[10,20]$ & $ 0.96\pm 0.10\pm 0.08$ & $ 0.74\pm 0.16\pm 0.11$ & $ 0.93\pm 0.10\pm 0.08$ \\
$[20,30]$ & $ 1.36\pm 0.22\pm 0.11$ & $ 1.05\pm 0.34\pm 0.16$ & $ 1.27\pm 0.20\pm 0.11$ \\
$[30,40]$ & $ 1.17\pm 0.32\pm 0.09$ & $ 0.93\pm 0.57\pm 0.14$ & $ 1.18\pm 0.31\pm 0.10$ \\
$[40,50]$ & $ 1.38\pm 0.47\pm 0.11$ & $ 1.60\pm 0.85\pm 0.24$ & $ 1.28\pm 0.40\pm 0.11$ \\
$[50,100]$ & $ 0.79\pm 0.23\pm 0.06$ & $ 1.51\pm 1.08\pm 0.23$ & $ 0.89\pm 0.28\pm 0.07$ \\
\hline
Centrality    & $ $ & $ $ & $ $ \\ \hline
$[0,10]$\% & $ 1.04\pm 0.06\pm 0.07$ & $ 1.10\pm 0.12\pm 0.16$ & $ 1.10\pm 0.06\pm 0.07$ \\
$[10,20]$\% & $ 1.12\pm 0.08\pm 0.08$ & $ 1.01\pm 0.12\pm 0.15$ & $ 1.14\pm 0.08\pm 0.08$ \\
$[20,30]$\% & $ 0.99\pm 0.09\pm 0.08$ & $ 1.12\pm 0.17\pm 0.17$ & $ 1.12\pm 0.09\pm 0.09$ \\
$[30,40]$\% & $ 1.14\pm 0.12\pm 0.11$ & $ 0.83\pm 0.17\pm 0.13$ & $ 1.06\pm 0.11\pm 0.10$ \\
$[40,50]$\% & $ 0.91\pm 0.14\pm 0.11$ & $ 0.86\pm 0.25\pm 0.15$ & $ 0.94\pm 0.14\pm 0.11$ \\
$[50,100]$\% & $ 1.14\pm 0.17\pm 0.18$ & $ 0.78\pm 0.28\pm 0.16$ & $ 1.17\pm 0.17\pm 0.18$ \\ \hline
$[0,100]$\%   & $ 1.06\pm 0.05\pm 0.08$ & $ 1.02\pm 0.08\pm 0.15$ & $ 1.10\pm 0.05\pm 0.09$   \\
\end{tabular}
\end{table}

\subsection{Combined results for the two decay channels}

According to lepton universality and given the large mass, the Z boson is expected to decay into the dimuon and dielectron channels with branching ratios within 1\% of each other. Also, neither muons nor electrons are expected to interact strongly with the medium formed in the collision. The two channels can therefore be checked against each other, and used to measure the combined $\cPZ\to \ell^+\ell^-$ yields and $R_\mathrm{AA}$, where $\cPZ\to \ell^+\ell^-$ refers to the Z boson decaying into either the dimuon or dielectron channel. Given the uncertainties in the measurements, in the region of overlap, the datasets are in agreement. The combination is then done following the best linear unbiased estimate technique, as described in Ref.~\cite{Lyons1988110}.

\begin{figure}[ht]
  \begin{center}
  \includegraphics[width=0.45\textwidth]{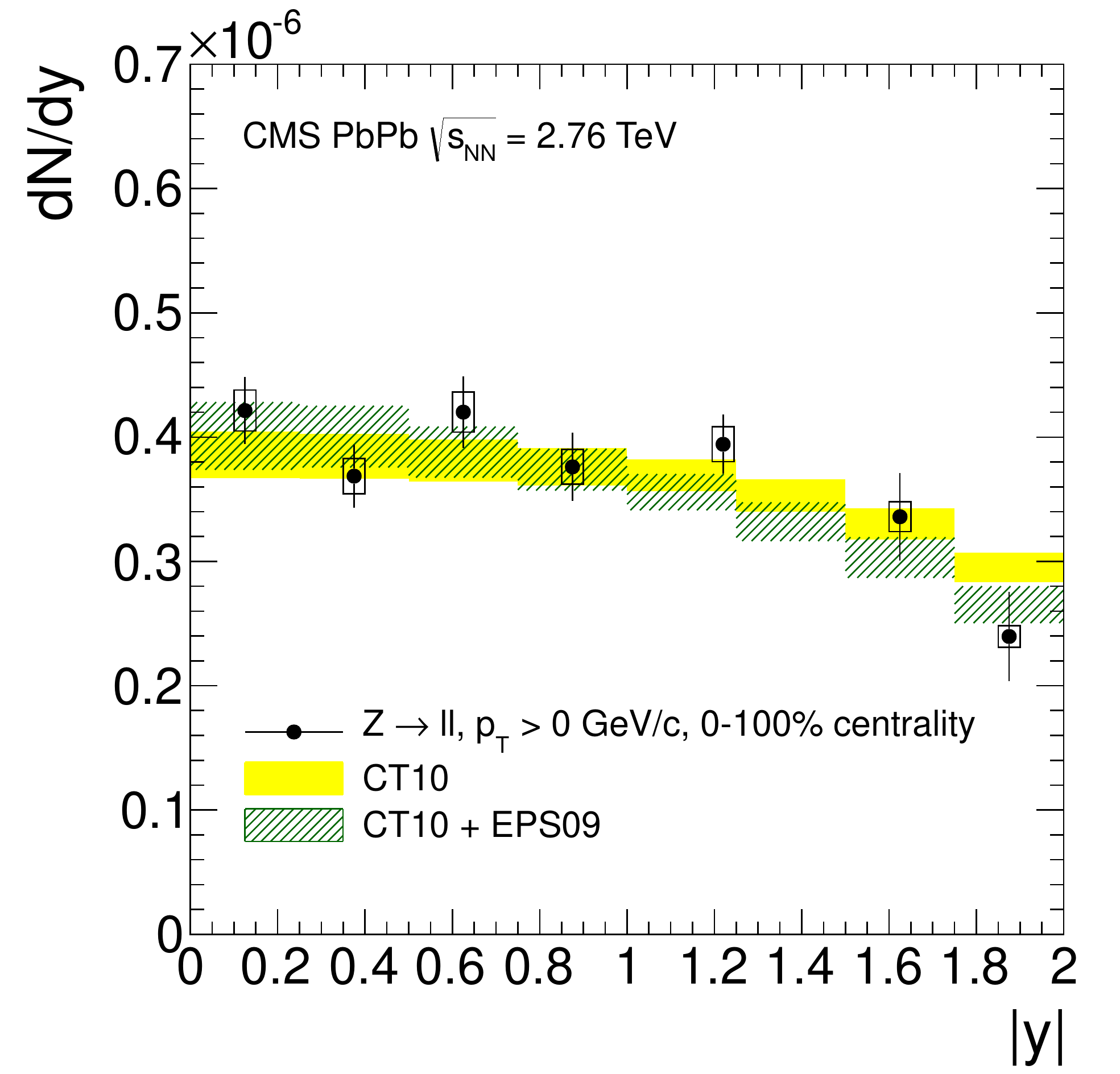}
  \includegraphics[width=0.45\textwidth]{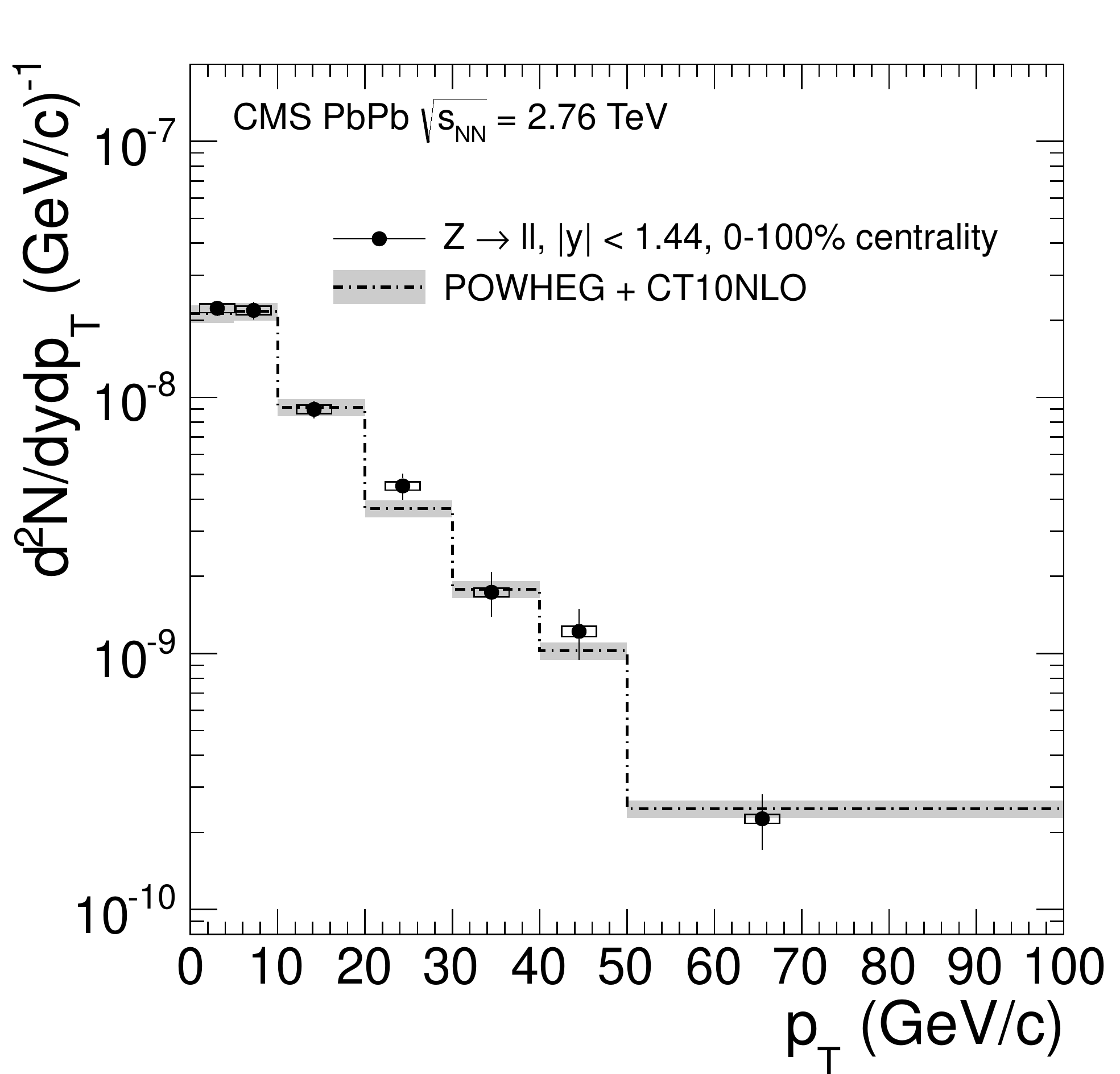}
  \includegraphics[width=0.45\textwidth]{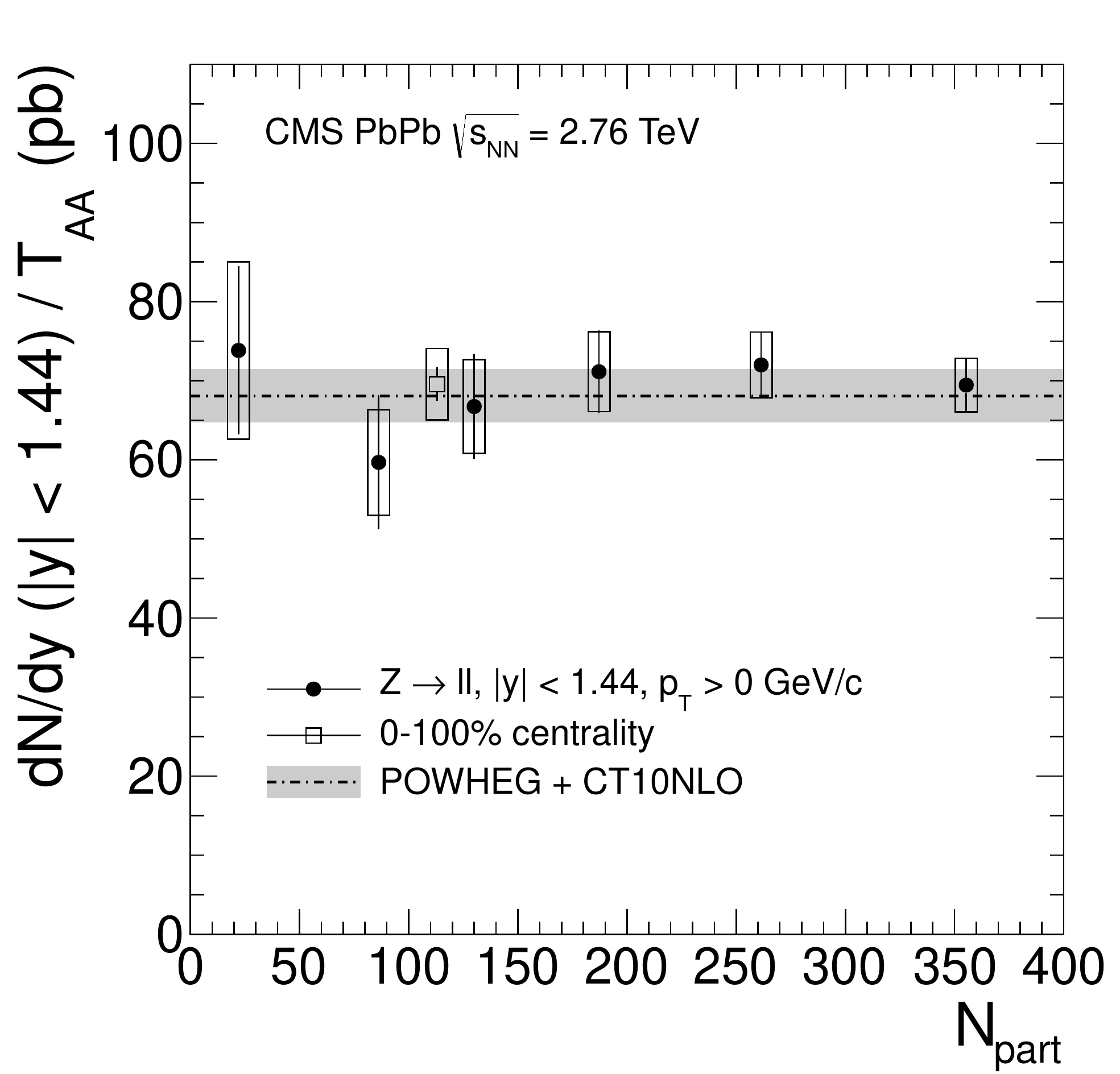}
  \caption{The measured $\cPZ\to \ell^+\ell^-$ yields per MB event in PbPb collisions, shown for the combined leptonic channel as a function of the Z boson $y$ (top left), \pt (top right), and $N_\text{part}$ (bottom). For the $y$ dependence, the measurements from dimuons and dielectrons are combined for $\abs{y} < 1.44$, and the dimuon measurements alone are shown for $1.5 < \abs{y}< 2.0$. The yields are compared with $\Pp\Pp\to\cPZ\to \ell^+\ell^-$ \POWHEG predictions scaled by the 0--100\% centrality averaged $T_\mathrm{AA}$. The light gray bands represent the theoretical uncertainty of 5\% assumed for the \POWHEG reference curve together with the uncertainty of 6.2\% due to the $T_\mathrm{AA}$ scaling. The results vs. $y$ are compared to predictions with (green dark band) and without (yellow light band) nuclear modification effects. Vertical lines (boxes) correspond to statistical (systematic) uncertainties.}
  \label{fig:YieldCombined}
  \end{center}
\end{figure}

\begin{figure}[ht]
  \begin{center}
  \includegraphics[width=0.45\textwidth]{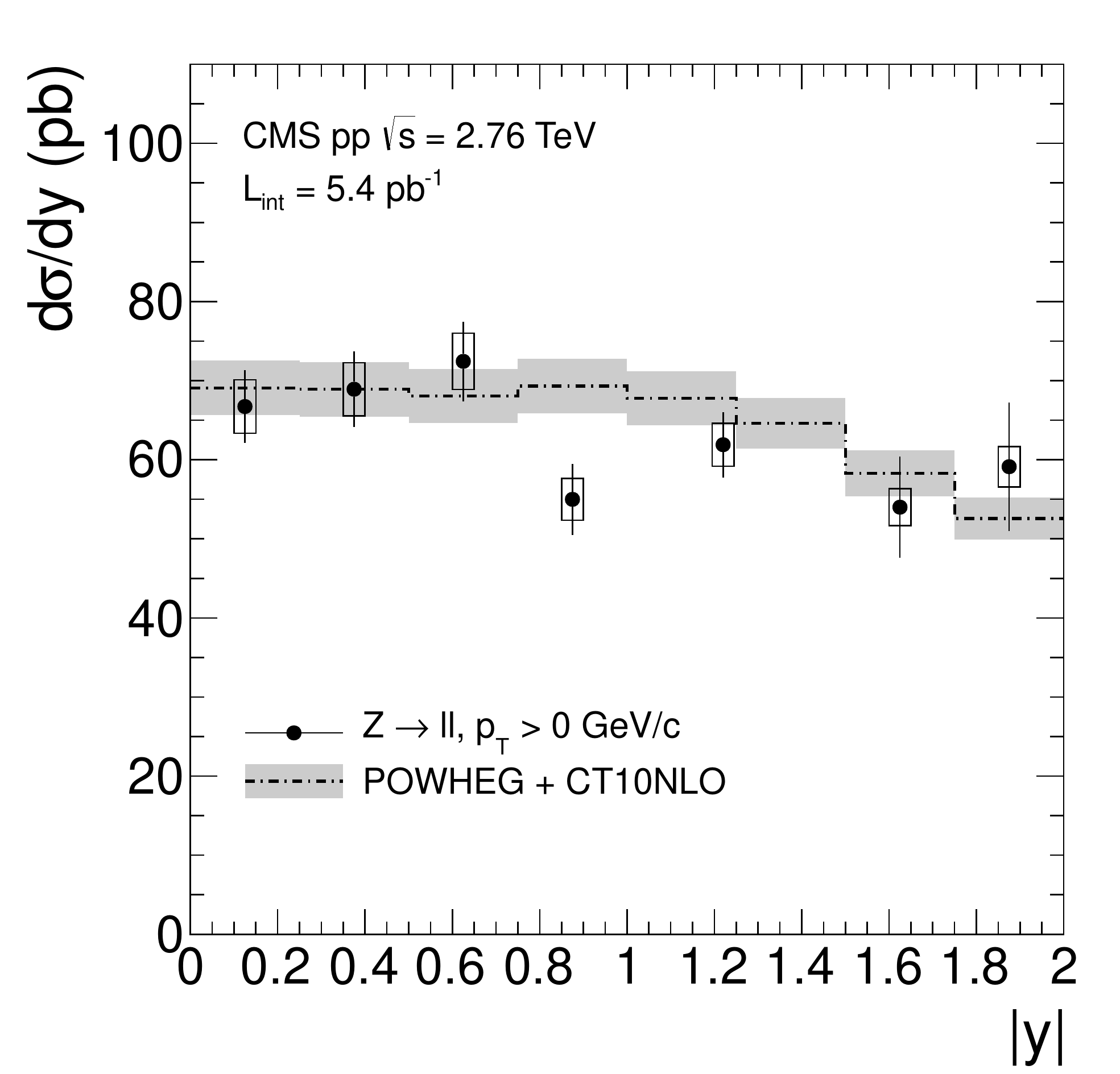}
  \includegraphics[width=0.45\textwidth]{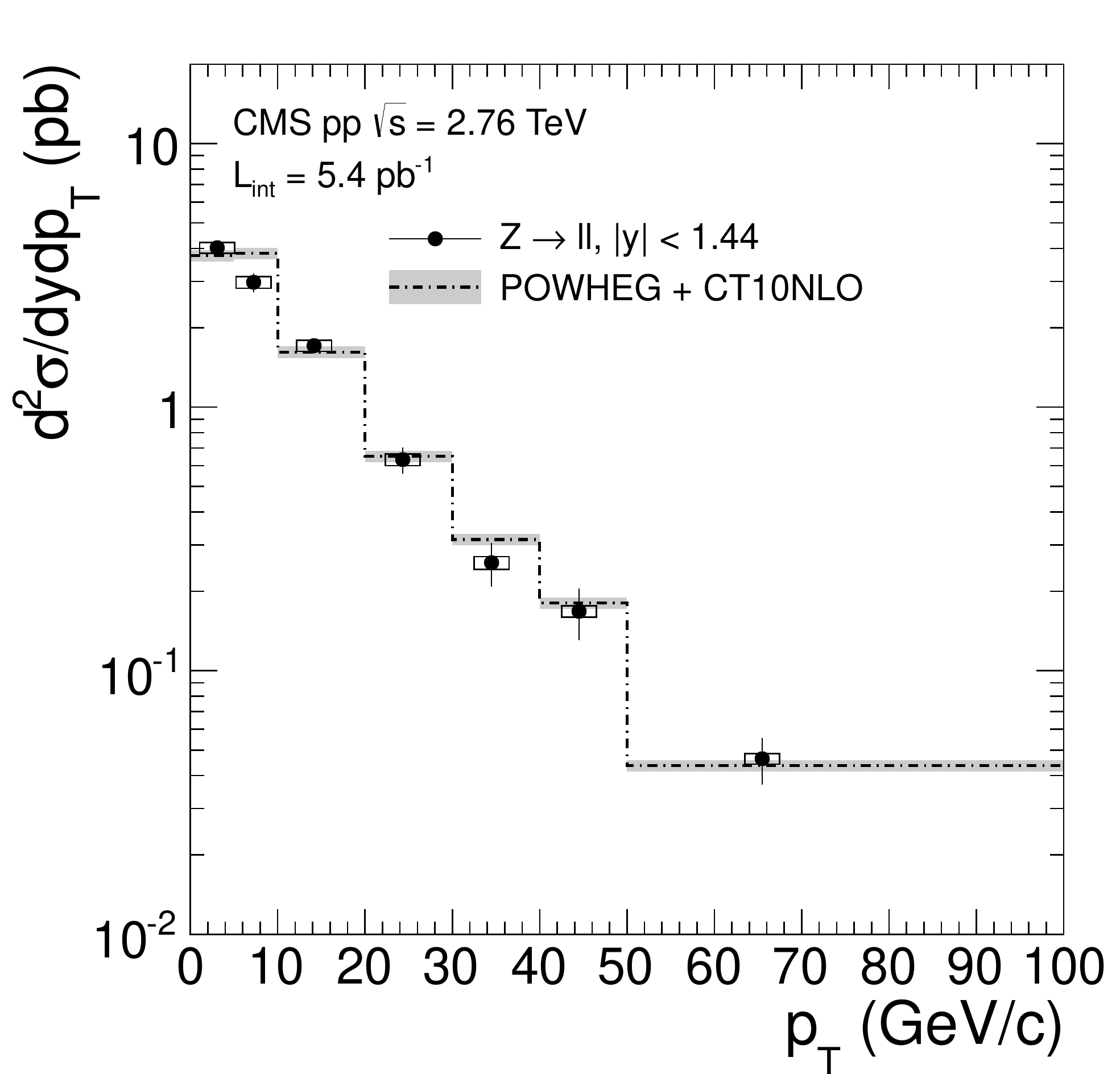}
  \caption{The measured $\cPZ\to \ell^+\ell^-$ cross section in pp collisions, shown for the combined leptonic channel as a function of the Z boson $y$ (left) and \pt (right). For the $y$ dependence, the measurements from dimuons and dielectrons are combined for $\abs{y} < 1.44$, and the dimuon measurements alone are shown for $1.5 < \abs{y}< 2.0$. Results are compared with $\Pp\Pp\to\cPZ\to \ell^+\ell^-$ \POWHEG predictions. Vertical lines (boxes) correspond to statistical (systematic) uncertainties. The theoretical uncertainty of 5\% assumed for the \POWHEG reference curve is shown by the grey band.}
  \label{fig:YieldCombinedpp}
  \end{center}
\end{figure}

The combined yields per MB event for PbPb collisions and the combined cross sections for pp collisions are shown in Figs.~\ref{fig:YieldCombined} and~\ref{fig:YieldCombinedpp}, respectively. The dimuon and dielectron measurements share the kinematic region of $\abs{y} < 1.44$. The dependence on \pt and $N_\text{part}$ of the Z boson yield and $R_\mathrm{AA}$ measurements in the combination of the two channels are therefore restricted to $\abs{y} < 1.44$. The dependence on $\abs{y}$ is shown with the combined measurements for $\abs{y} < 1.44$, extended with the dimuon measurements for the $1.5 < \abs{y} < 2.0$ range.

The results as a function of \pt, $y$, and centrality are compared with predictions from the \POWHEG generator; this comparison shows that the measurements agree with the theoretical calculations within the combined statistical and systematic uncertainties. The current precision of the measurements does not allow to distinguish between the unbound proton PDF sets and the modified nuclear PDF sets.

\begin{figure}[ht]
  \begin{center}
  \includegraphics[width=0.45\textwidth]{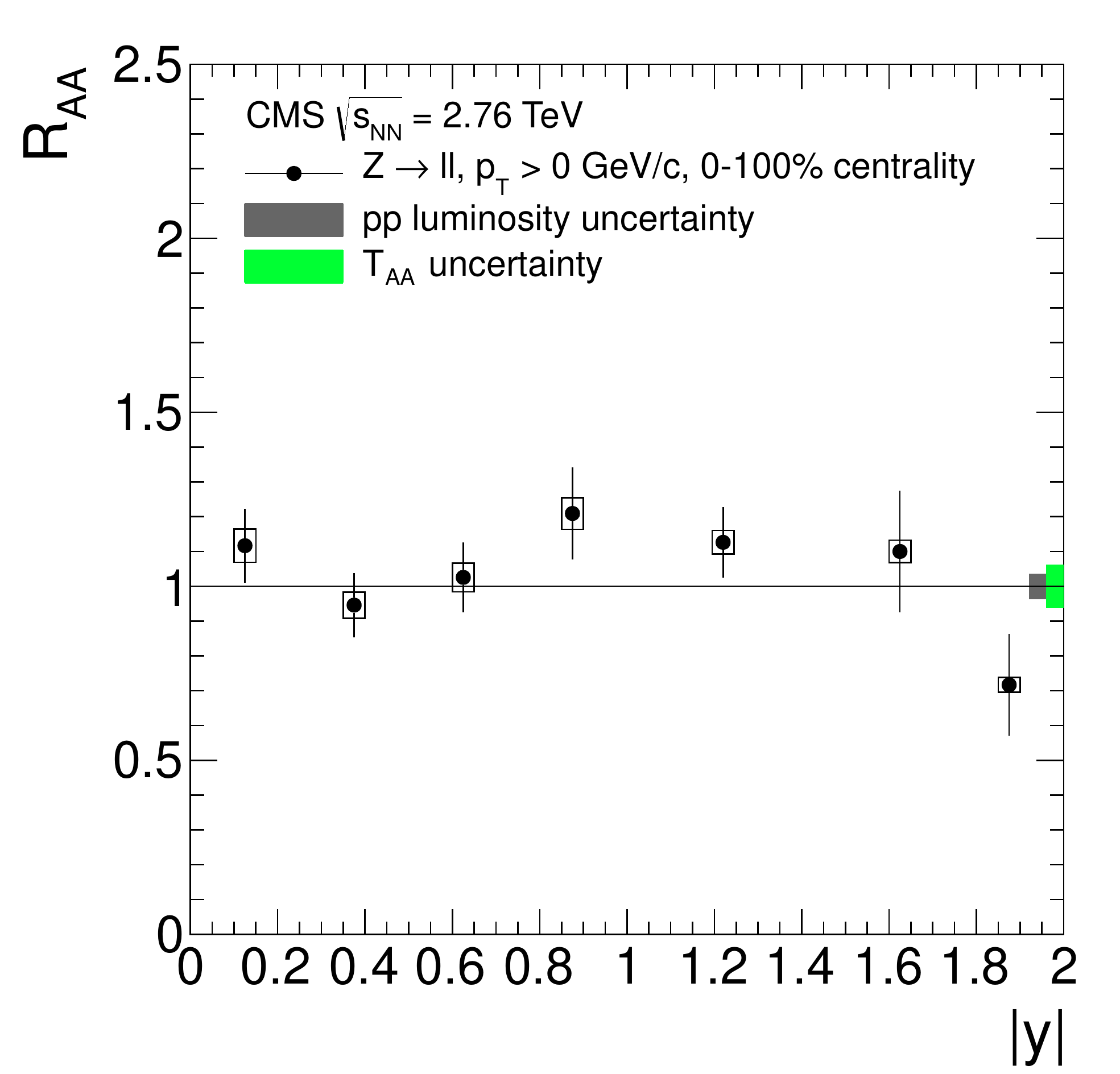}
  \includegraphics[width=0.45\textwidth]{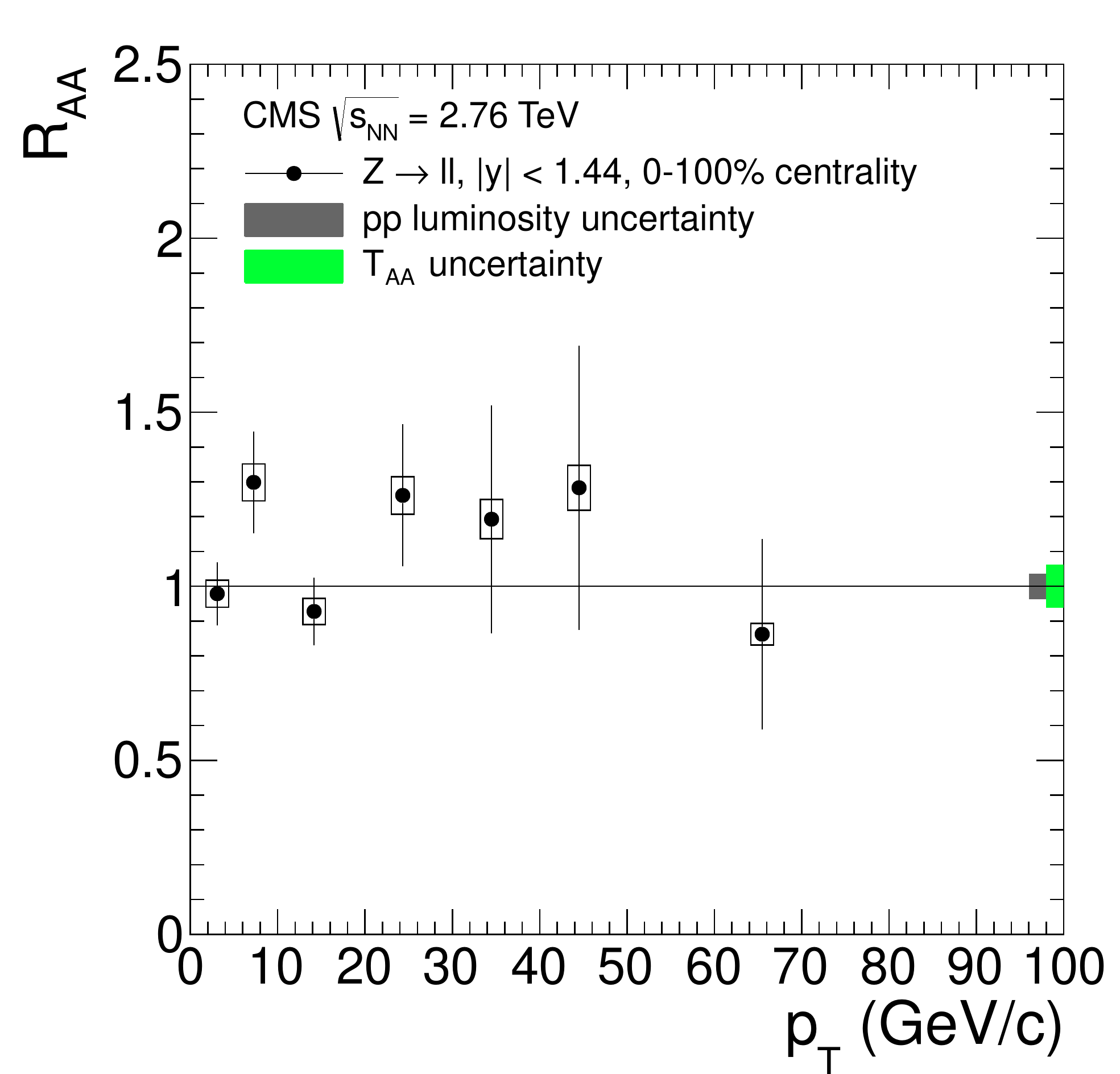}
  \includegraphics[width=0.45\textwidth]{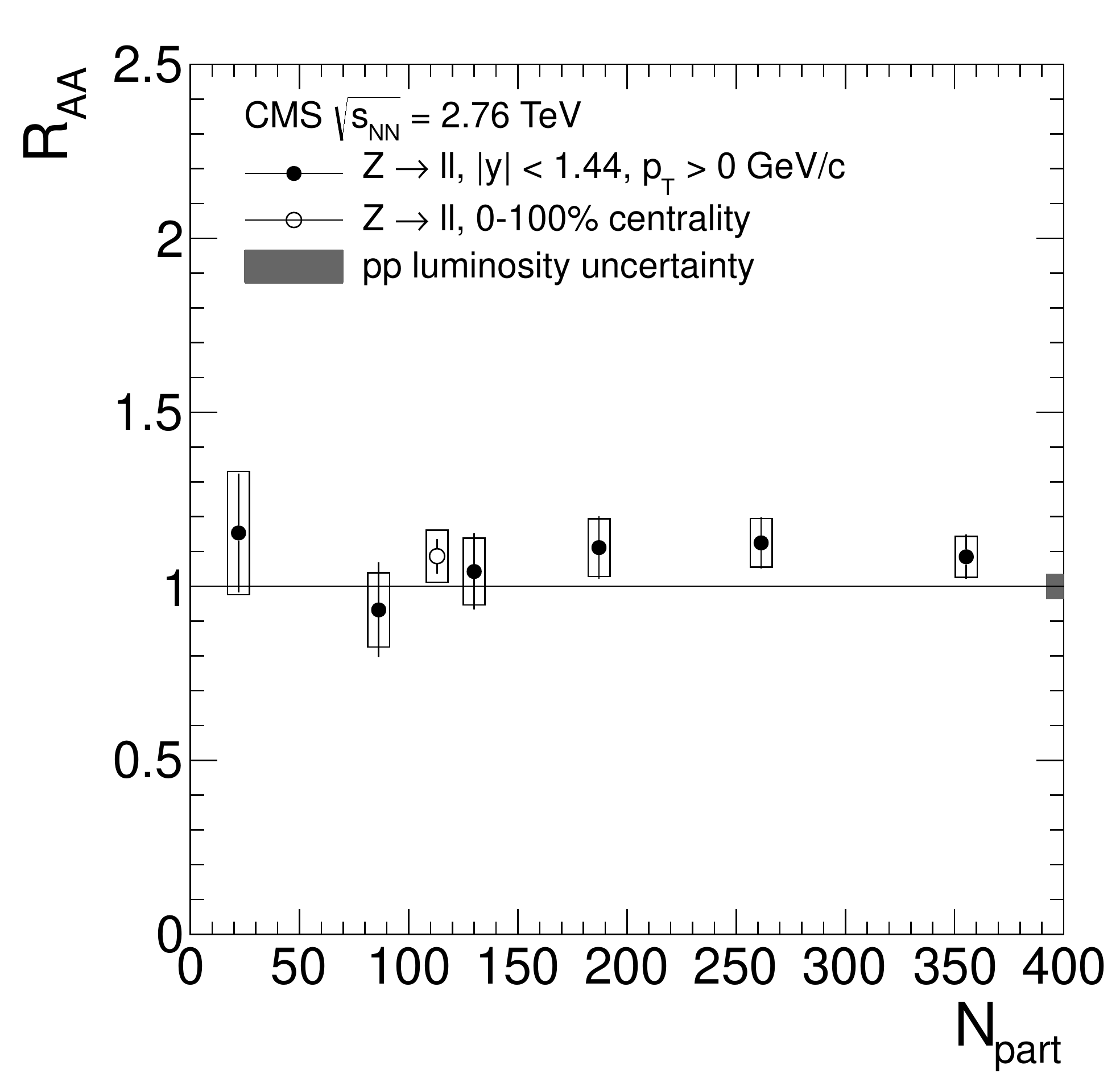}
  \caption{The Z boson $R_\mathrm{AA}$ values for the combination of the dimuon and dielectron channel, as a function of $y$ (top left), \pt (top right), and $N_\text{part}$ (bottom). The horizontal line at $R_\mathrm{AA}$ = 1 is drawn as a reference. Vertical lines (boxes) correspond to statistical (systematic) uncertainties. Grey light bar at $R_\mathrm{AA} = 1$ corresponds to uncertainty in pp luminosity and green dark bar corresponds to uncertainty of $T_\mathrm{AA}$.}
  \label{fig:RAACombined}
  \end{center}
\end{figure}

To calculate the combined $R_\mathrm{AA}$, the combined dilepton yields in PbPb and pp data are obtained and then the $R_\mathrm{AA}$ ratio is calculated based on those values. The combined $R_\mathrm{AA}$ values are given in Table~\ref{tab:all_together} and in Fig.~\ref{fig:RAACombined}. The $R_\mathrm{AA}$ for the combination of the two channels shows no dependence and no variation in nuclear effects as a function of \pt, $y$, or centrality. This demonstrates that within uncertainties, Z boson production is not modified in PbPb collisions compared with pp collisions scaled by the number of inelastic nucleon-nucleon collisions.

\section{Summary}

The yields of Z bosons have been measured as a function of \pt, $y$, and centrality, in both the dimuon and dielectron channels for PbPb collisions at $\sqrtsnn$ of 2.76\TeV with an integrated luminosity of approximately $166$\mubinv. The $\cPZ\to \Pgmp\Pgmm$ and $\cPZ\to \Pep\Pem$ cross sections have been measured in pp collisions at the same collision energy with an integrated luminosity of 5.4\pbinv. Within the combined statistical and systematic uncertainties, no centrality dependence is observed once the yields are normalized by the number of inelastic nucleon-nucleon collisions. When integrated over centrality, the Z boson $y$ and \pt distributions are found to be consistent between the PbPb and pp data and also to agree with theoretical predictions. The centrality-integrated $R_\mathrm{AA}$ is found to be $1.06\pm0.05\stat\pm0.08\syst$ in the dimuon channel and $1.02\pm0.08\stat\pm0.15\syst$ in the dielectron channel. No significant nuclear modifications are found as a function of \pt, $y$, or centrality in either the dimuon or dielectron channels over the entire kinematic range studied.

\begin{acknowledgments}
We congratulate our colleagues in the CERN accelerator departments for the excellent performance of the LHC and thank the technical and administrative staffs at CERN and at other CMS institutes for their contributions to the success of the CMS effort. In addition, we gratefully acknowledge the computing centres and personnel of the Worldwide LHC Computing Grid for delivering so effectively the computing infrastructure essential to our analyses. Finally, we acknowledge the enduring support for the construction and operation of the LHC and the CMS detector provided by the following funding agencies: BMWFW and FWF (Austria); FNRS and FWO (Belgium); CNPq, CAPES, FAPERJ, and FAPESP (Brazil); MES (Bulgaria); CERN; CAS, MoST, and NSFC (China); COLCIENCIAS (Colombia); MSES and CSF (Croatia); RPF (Cyprus); MoER, ERC IUT and ERDF (Estonia); Academy of Finland, MEC, and HIP (Finland); CEA and CNRS/IN2P3 (France); BMBF, DFG, and HGF (Germany); GSRT (Greece); OTKA and NIH (Hungary); DAE and DST (India); IPM (Iran); SFI (Ireland); INFN (Italy); NRF and WCU (Republic of Korea); LAS (Lithuania); MOE and UM (Malaysia); CINVESTAV, CONACYT, SEP, and UASLP-FAI (Mexico); MBIE (New Zealand); PAEC (Pakistan); MSHE and NSC (Poland); FCT (Portugal); JINR (Dubna); MON, RosAtom, RAS and RFBR (Russia); MESTD (Serbia); SEIDI and CPAN (Spain); Swiss Funding Agencies (Switzerland); MST (Taipei); ThEPCenter, IPST, STAR and NSTDA (Thailand); TUBITAK and TAEK (Turkey); NASU and SFFR (Ukraine); STFC (United Kingdom); DOE and NSF (USA).

Individuals have received support from the Marie-Curie programme and the European Research Council and EPLANET (European Union); the Leventis Foundation; the A. P. Sloan Foundation; the Alexander von Humboldt Foundation; the Belgian Federal Science Policy Office; the Fonds pour la Formation \`a la Recherche dans l'Industrie et dans l'Agriculture (FRIA-Belgium); the Agentschap voor Innovatie door Wetenschap en Technologie (IWT-Belgium); the Ministry of Education, Youth and Sports (MEYS) of the Czech Republic; the Council of Science and Industrial Research, India; the HOMING PLUS programme of Foundation for Polish Science, cofinanced from European Union, Regional Development Fund; the Compagnia di San Paolo (Torino); the Consorzio per la Fisica (Trieste); MIUR project 20108T4XTM (Italy); the Thalis and Aristeia programmes cofinanced by EU-ESF and the Greek NSRF; and the National Priorities Research Program by Qatar National Research Fund.
\end{acknowledgments}
\bibliography{auto_generated}   
\cleardoublepage \appendix\section{The CMS Collaboration \label{app:collab}}\begin{sloppypar}\hyphenpenalty=5000\widowpenalty=500\clubpenalty=5000\textbf{Yerevan Physics Institute,  Yerevan,  Armenia}\\*[0pt]
S.~Chatrchyan, V.~Khachatryan, A.M.~Sirunyan
\vskip\cmsinstskip
\textbf{Institut f\"{u}r Hochenergiephysik der OeAW,  Wien,  Austria}\\*[0pt]
W.~Adam, T.~Bergauer, M.~Dragicevic, J.~Er\"{o}, C.~Fabjan\cmsAuthorMark{1}, M.~Friedl, R.~Fr\"{u}hwirth\cmsAuthorMark{1}, V.M.~Ghete, C.~Hartl, N.~H\"{o}rmann, J.~Hrubec, M.~Jeitler\cmsAuthorMark{1}, W.~Kiesenhofer, V.~Kn\"{u}nz, M.~Krammer\cmsAuthorMark{1}, I.~Kr\"{a}tschmer, D.~Liko, I.~Mikulec, D.~Rabady\cmsAuthorMark{2}, B.~Rahbaran, H.~Rohringer, R.~Sch\"{o}fbeck, J.~Strauss, A.~Taurok, W.~Treberer-Treberspurg, W.~Waltenberger, C.-E.~Wulz\cmsAuthorMark{1}
% ...
\vskip\cmsinstskip
% ...
1:~~Also at Vienna University of Technology, Vienna, Austria\\
2:~~Also at CERN, European Organization for Nuclear Research, Geneva, Switzerland\\
% ...

\end{sloppypar}
\end{document}